\documentclass[aip,rsi,amsmath,amssymb,reprint]{revtex4-1}

\usepackage[normalem]{ulem} 
\usepackage{amsmath,bm}
\usepackage{amsfonts}
\usepackage{amssymb,enumerate}
\usepackage{graphicx,hyperref}
\usepackage{amssymb}
\usepackage{epstopdf,booktabs}
\usepackage{empheq}
\usepackage{fancybox}
\usepackage{subcaption}

\usepackage[usenames,dvipsnames]{color}
\usepackage[T1]{fontenc}
\usepackage[utf8]{inputenc}
\usepackage{natbib}
\usepackage{soul}

\usepackage{verbatim}
\usepackage{xcolor}
\definecolor{beaublue}{rgb}{0.74, 0.83, 0.9}
\definecolor{gainsboro}{rgb}{0.86, 0.86, 0.86}
\definecolor{lavender(web)}{rgb}{0.9, 0.9, 0.98}

\definecolor{shadowcolor}{rgb}{0,.5,.5}
\numberwithin{equation}{section}

\newcommand{\be}{\begin {align}}
\newcommand{\ee}{\end {align}}
\newcommand{\beqa}{\begin {eqnarray}}
\newcommand{\eeqa}{\end {eqnarray}}

\hypersetup{
    colorlinks,    citecolor=blue,    filecolor=blue,    linkcolor=blue,    urlcolor=blue
}

\begin{document}

\begin{abstract}

We demonstrate that laser reflection acts as a catalyst for superponderomotive electron production in the preplasma formed by relativistic multipicosecond lasers incident on solid density targets.
In 1D particle-in-cell simulations, high energy electron production proceeds via two stages of direct laser acceleration, an initial stochastic backward stage, and a final non-stochastic forward stage.
The initial stochastic stage, driven by the reflected laser pulse, provides the pre-acceleration needed to enable the final stage to be non-stochastic.
Energy gain in the electrostatic potential, which has been frequently considered to enhance stochastic heating, is only of secondary importance.
The mechanism underlying the production of high energy electrons by laser pulses incident on solid density targets is of direct relevance to applications involving multipicosecond laser-plasma interactions.

\end{abstract}

\title{Laser reflection as a catalyst for direct laser acceleration in multipicosecond laser-plasma interaction}

\author{K. Weichman}
\affiliation{Department of Mechanical and Aerospace Engineering, University of California at San Diego, La Jolla, CA 92093, USA}
\author{A.P.L. Robinson}
\affiliation{Central Laser Facility, STFC Rutherford-Appleton Laboratory, Didcot, OX11 0QX, UK}
\author{F.N. Beg}
\affiliation{Department of Mechanical and Aerospace Engineering, University of California at San Diego, La Jolla, CA 92093, USA}
\author{A.V. Arefiev}
\affiliation{Department of Mechanical and Aerospace Engineering, University of California at San Diego, La Jolla, CA 92093, USA}

\date{\today}

\maketitle


\section{Introduction}

Laser-plasma interaction (LPI) at relativistic intensities ($>10^{18}$ W/cm$^2$) provides an efficient, compact source of energetic charged particles, neutrons, and radiation useful for applications in accelerator science \cite{tajima1979,malka2012accelerators_review}, 
imaging \cite{li2006proton_radiography,kneip2011frog}, 
laboratory astrophysics \cite{bulanov2015labastro}, 
and inertial confinement fusion \cite{lindl1995icf}.
Often, the production of accelerated ions \cite{roth2001proton_icf,macchi2013ion_accel_rev}, 
neutrons \cite{pomerantz2014neutrons},
or high energy X-rays \cite{kmetec1992brem,stark2016xray}
is driven indirectly by the laser through the production of high energy electrons.
Understanding the mechanisms of electron heating via LPI is therefore crucial for a variety of applications.

In low density plasma, there are several routes for electron acceleration, including direct laser acceleration (DLA) \cite{hartemann1995dla}
and wakefield acceleration \cite{tajima1979laser}, 
capable of producing electrons with energies of MeV or greater.
A laser pulse incident on a solid density plasma reflects from the surface where the electron density $n_e \approx \gamma_0 n_{cr}$ (the critical density $n_{cr} = \omega^2 m_e/4\pi e^2$ is adjusted by the relativistic factor $\gamma_0 \approx (1+a_0^2)^{1/2}$ associated with transverse oscillations in the laser, which has normalized amplitude $a_0 = eE_0/m_e c \omega$ \cite{kaw1970rel_trans}),
producing an electric field that can be interpreted as a standing wave close to the surface and counter-propagating pulses further away.
Energy gain mechanisms at a sharp laser-solid interface for a laser pulse at normal incidence include j$\times$B heating \cite{kruer1985jxB} and 
standing wave acceleration \cite{lefebvre1997absorption,may2011absorption}, and typically generate peak electron energy near the ponderomotive limit $E_p = m_e c^2(1+ a_0^2/2)$ and a slope temperature for the electron energy distribution consistent with the ponderomotive scaling $T_p = [(1+a_0^2)^{1/2}-1]m_e c^2$ \cite{wilks1992pond_scale}.

When the interface between the vacuum and the opaque plasma is not sharp, superponderomotive electron acceleration in excess of tens to hundreds of MeV with $a_0 < 10$ is possible via a variety of mechanisms \cite{pukhov1999betatron,arefiev2012channel_accel,kemp2009psshelf_pond}.
In particular, the generation of high energy electrons in the presence of the counter-propagating incident and reflected laser components produced by incomplete laser absorption has generated significant interest in stochastic heating \cite{mendonca1983stochastic,rax1992stochastic_heating} as a source of highly energetic, superponderomotive ($\gamma \gg 1+a_0^2/2$) electrons in long scale-length plasma \cite{sheng2002stochastic_theory,kemp2009psshelf_pond,paradkar2011scalelength,kojima2019electromagnetic}.

The development of a large region of low density plasma in front of an opaque target is virtually unavoidable for relativistically intense pulses with picosecond duration. 
Picosecond pulses incident on solid-density targets evolve an initial exponential preplasma density distribution towards a long, relatively flat subcritical plateau jumping up sharply to overcritical density \cite{kemp2009psshelf_pond,paradkar2011scalelength,sorokovikova2016superpond,iwata2018hole_boring}.
The preplasma produced in this way can have a quasi-1D geometry when the laser spot size is large, as is available from existing high power laser facilities such as LFEX \cite{Miyanaga2006LFEX}, LMJ-PETAL 
\cite{batani2014PETAL}, NIF ARC \cite{crane2010NIF_ARC}, and OMEGA EP \cite{maywar2008OMEGA_EP}.
While higher-dimensional effects can somewhat alter the electron and ion dynamics, for instance reducing the reflectively of the opaque plasma surface \cite{kemp2012ps_2d,iwata2018hole_boring},
1D simulations present a useful platform for developing a conceptual understanding of picosecond laser-solid interaction under experimentally relevant conditions.

While substantial progress has been made in understanding possible routes to energetic electron production by picosecond pulses, the electron acceleration process observed in portions of the parameter space remains to be fully explained.
Concurrent with the development of the characteristic preplasma "shelf" and solid density "wall" plasma profile in Ref. \onlinecite{sorokovikova2016superpond} was the observation that high energy electrons originate in the wall and follow a trajectory consisting of initial backward propagation along the shelf, followed by reflection to forward propagation, during which strong DLA occurs.
It can be deduced from the dephasing rate and the laser work done on the characteristic high energy electrons shown in Ref. \onlinecite{sorokovikova2016superpond} (Appendix \ref{append:vphi} includes a demonstration of this analysis) that high energy electrons were accelerated \textit{without} slipping substantially in the phase they experience in the forward-propagating component of the laser, 
which is inconsistent with the high phase slip characteristic of stochastic heating.
While it has been suggested that a sufficient electrostatic potential well in the ion shelf region may serve to "lock" the electron in phase with the co-propagating laser, enhancing electron energy in a multi-bounce process \cite{paradkar2011scalelength},
the energy gain observed in Ref. \onlinecite{sorokovikova2016superpond} occurs with only a \textit{single} bounce and is \textit{substantially in excess} of the well depth (the maximum beam energy produced by this mechanism for a finite potential well \cite{paradkar2012wellheating_theory}).

The purpose of the present work is to demonstrate that the reflected component of the laser pulse enables non-stochastic electron acceleration to high energy along a single bounce trajectory in the ion shelf-wall density profile characteristic of picosecond laser-solid interactions.
For $a_0 = 5$ and an underdense plasma shelf of 50 $\mu$m, we observe electron acceleration via direct laser acceleration in excess of 75 MeV, more than 10 times the vacuum ponderomotive limit.
It will be shown that the production of such high energy electrons during the forward DLA
requires accounting for electron energization by the backward-propagating reflected component of the laser pulse and that this pre-acceleration enables non-stochastic energy gain in excess of what can be produced by stochastic heating.
In contrast with previous work where the electron energy gain was found to correspond to stochastic heating or the electrostatic potential played an important role in reducing phase slip, we find that a DLA-plus-bounce model in which the only contribution of the electrostatic potential is to reflect energetic electrons is sufficient to reproduce the observed pre-acceleration and final non-stochastic energy gain.

The outline of this paper is as follows. Section \ref{sec:setup} introduces a simulation in which we observe strong electron acceleration by DLA processes which exhibit different character during backward and forward propagation and do not appear to require energy gain from the electrostatic potential (Section \ref{sec:sim-results}). In Section \ref{sec:DLA-model}, we demonstrate that DLA in counter-propagating pulses can produce the observed behavior and we identify the backward and forward acceleration processes as corresponding to stochastic and non-stochastic DLA, respectively, explaining the vital role of laser reflection in high energy electron production.

\section{Electron acceleration model}
\label{sec:setup}

\begin{figure*}
    \centering
    \begin{subfigure}{0.45\textwidth} 
    \includegraphics[width=0.95\linewidth]{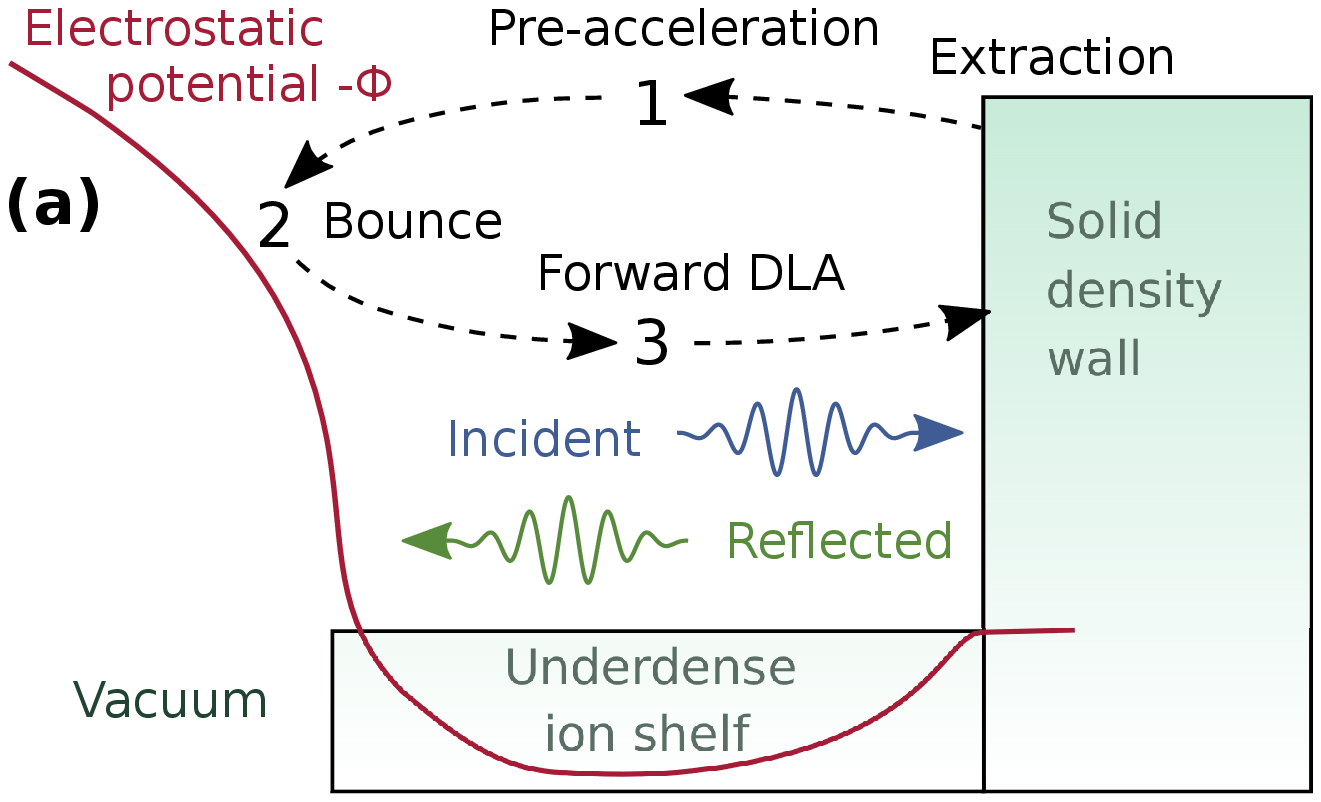}
    \subcaption{} \label{fig:cartoon}
    \end{subfigure}
    \begin{subfigure}{0.45\textwidth} 
    \includegraphics[width=0.95\linewidth]{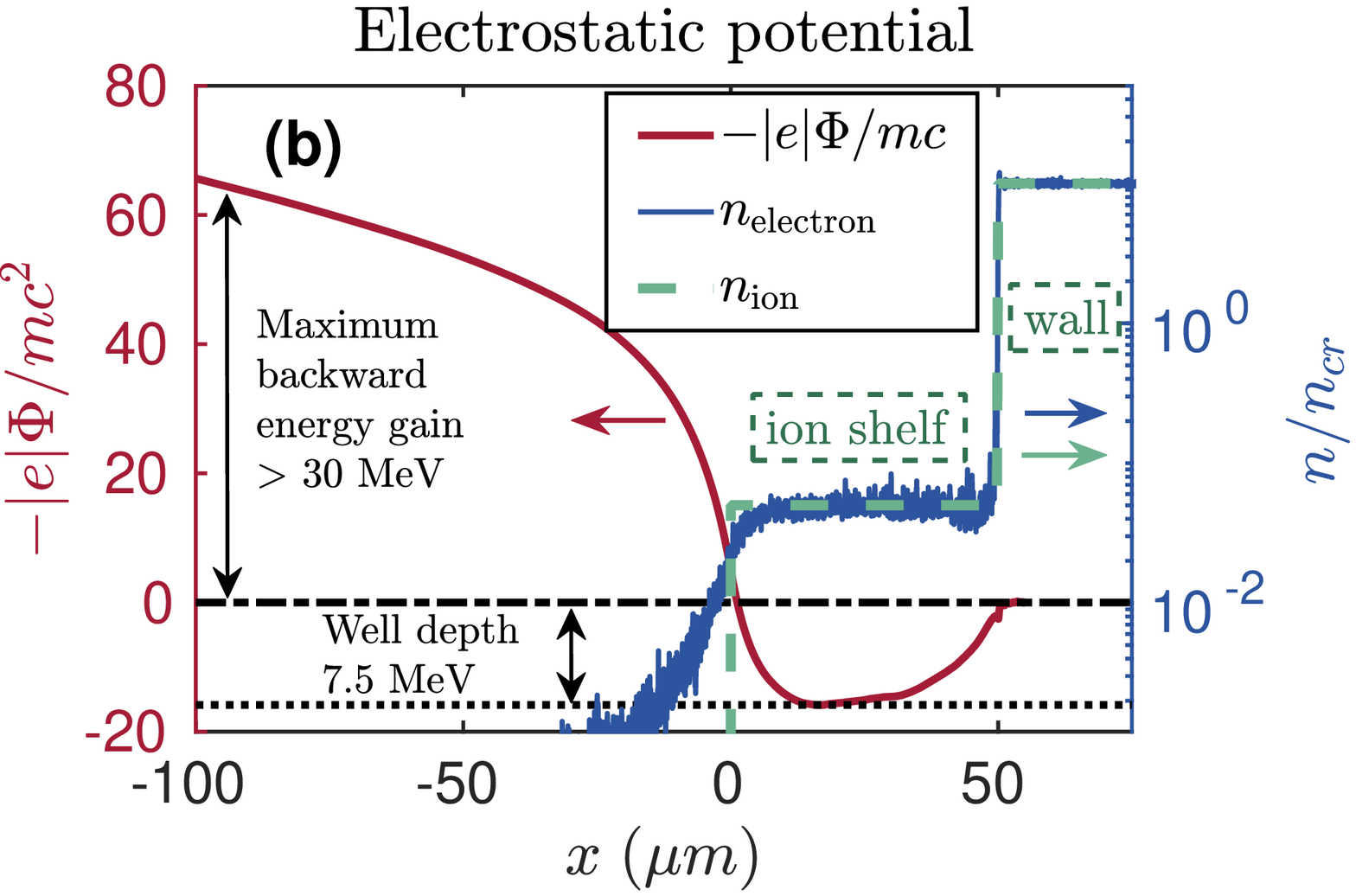}
    \subcaption{} \label{fig:phi_es}
    \end{subfigure}
    \vspace{-1.75\baselineskip}
    \caption{PIC simulation setup and electrostatic potential. (a) Conceptual diagram of simulation setup and electron acceleration process. 
    (b) Time-averaged electron density and electrostatic potential relative to the wall surface, with (immobile) ion density, from 1D PIC simulation.}
    \label{fig:cart-pot}
\end{figure*}

We model electron acceleration in the characteristic subcritical plateau ("shelf") and overcritical density ("wall") preplasma profile formed by multipicosecond pulses in 1D using the EPOCH paricle-in-cell (PIC) code \cite{arber2015epoch}. 
The preplasma consists of a 50 $\mu$m long 5\% critical density ($0.05 n_{cr}$) shelf in front of a solid density plasma wall modeled by 10 times the critical density, as shown conceptually in Figure \ref{fig:cartoon} and numerically in Figure \ref{fig:phi_es}. 
The wall extends to the far boundary of the simulation domain, which is sized such that a light speed object originating at the vacuum boundary of the simulation and being reflected from the far boundary could not make a second transit of the wall surface during the simulation. 
The vacuum boundary is located 100 $\mu$m from the shelf edge. 
Into this system is introduced a semi-infinite laser pulse with a rapid 100 fs HWHM rise time, which is short compared to the time it takes electrons to be accelerated.
The electric field of the laser is linearly polarized in $y$ and propagates in the $x$-direction.The simulation is allowed to run for 2.2 ps after the peak of the laser pulse hits the solid density plasma, which we find is sufficient for the electron energy spectrum to saturate.

The resolution of the simulation is 150 cells/$\mu$m, which gives 7.5 cells per skin depth ($c/\omega_{pe}$) in the solid density plasma. This resolution ensures direct laser acceleration is well-resolved \cite{arefiev2015resolution}.
The plasma is modeled by 3000 macroparticles per cell in both the shelf and the wall, for both electrons and ions. The ions are singly charged and treated as immobile.

The choice of a semi-infinite laser pulse and immobile ions allows us to study the electron dynamics occurring near the peak of a multipicosecond laser pulse. The semi-infinite pulse duration is appropriate for studying the electron acceleration process in multipicosecond pulses provided electrons are accelerated on a sub-ps timescale, which we find to be the case (discussed in Sections \ref{sec:sim-results} and \ref{sec:DLA-model}).
Using immobile ions allows us to isolate the effect of the presence of the shelf and wall ion density profile that arises in self-consistent multipicosecond simulations on electron dynamics from effects related to the time evolution of this profile, such as its initial establishment.
As we will demonstrate in Section \ref{sec:discuss}, the key conclusions of our work remain valid with mobile ions, even when the ion density evolves on the picosecond scale.
A 1D simulation with immobile ions may also artificially increase the plasma reflectivity at the critical density surface and  thereby may artificially increase the amplitude of the reflected wave compared to 2D simulations \cite{kemp2012ps_2d}. 
The intent of this work is to describe the energy gain mechanism we observe in the 1D simulations and provide a framework for evaluating energy gain in higher-dimensional simulations. As we will discuss in Section \ref{sec:discuss}, the mechanism we observe in our 1D simulations is still conceptually valid in systems with lower reflectivity.

We capture the time history of the position, momentum, and other properties of the individual (macro)electrons extracted from the wall and accelerated over the shelf region. The collected information includes a diagnostic for the longitudinal and transverse work done on each particle.
It is well-known that the energetic electrons produced by relativistic laser pulses interacting with plasma are predominantly forward-directed (momentum $|p_x| \gg |p_y|$, i.e. $p_y$ does not contribute substantially to the energy) despite the laser only doing transverse work ($W_y$) in 1D. 
We therefore use the work diagnostic to separate the contributions to the forward-directed momentum from the plasma-generated longitudinal electric field ($E_x$) and the laser electric field ($E_y$),
\begin{equation}
    W_{x,y} = - |e| \int_0^t dt' v_{x,y} E_{x,y}.
\end{equation}

\section{Direct laser acceleration in PIC simulation} 
\label{sec:sim-results}

We begin by outlining the results of our PIC simulation. 
Key features of the simulation setup and electron trajectory are illustrated in Figure \ref{fig:cartoon}. 
As we will show, the reflection of the incident (forward-propagating) laser pulse from the critical density surface is particularly important, and in 1D occurs with minimal absorption. 

Over the course of the simulation, the laser removes electrons from the immobile ion shelf and fosters the development of a substantial electrostatic (electron) potential well in the ion shelf region (Figure \ref{fig:phi_es}). The potential well develops over a few hundred fs and reaches a depth comparable to the ponderomotive potential. 
The development of a similarly deep potential well in picosecond laser-solid interaction is also observed in simulations with mobile ions \cite{kemp2009psshelf_pond,sorokovikova2016superpond} (see also Section \ref{sec:discuss}).
The maximum magnitude of the potential in the vacuum region, to the left of the ion shelf in Figure \ref{fig:phi_es}, is far in excess of the magnitude at the wall surface. This asymmetry in the potential substantially exceeds the initial kinetic energy of electrons entering the shelf region from the wall surface (the maximum value it could have in a purely electrostatic system) and indicates that electrons undergo substantial backward-directed energy gain from non-electrostatic sources.

The incident and reflected (backward-propagating) components of the laser accelerate electrons to high energies in a characteristic single-bounce trajectory (Figure \ref{fig:cartoon}).
The majority of high energy electrons originate in the solid density wall and either begin close to the surface or are brought to the surface as return current. 
The surface oscillates under the influence of the reflecting laser pulse \cite{wilks1993osc_surface,bulanov1994osc_surface} and ejects electrons into the shelf region.
Electrons extracted from the surface of the wall are then
(1) initially accelerated backwards through the underdense preplasma, (2) reflected off of the electrostatic potential in the vicinity of the shelf-vacuum interface, and (3) re-accelerated forwards through the ion shelf and re-injected into the overdense wall. 
Steps 1-3 take around 500 fs on average.

\begin{figure*}
    \centering
    \begin{subfigure}{0.47\textwidth} 
    \includegraphics[width=0.95\linewidth]{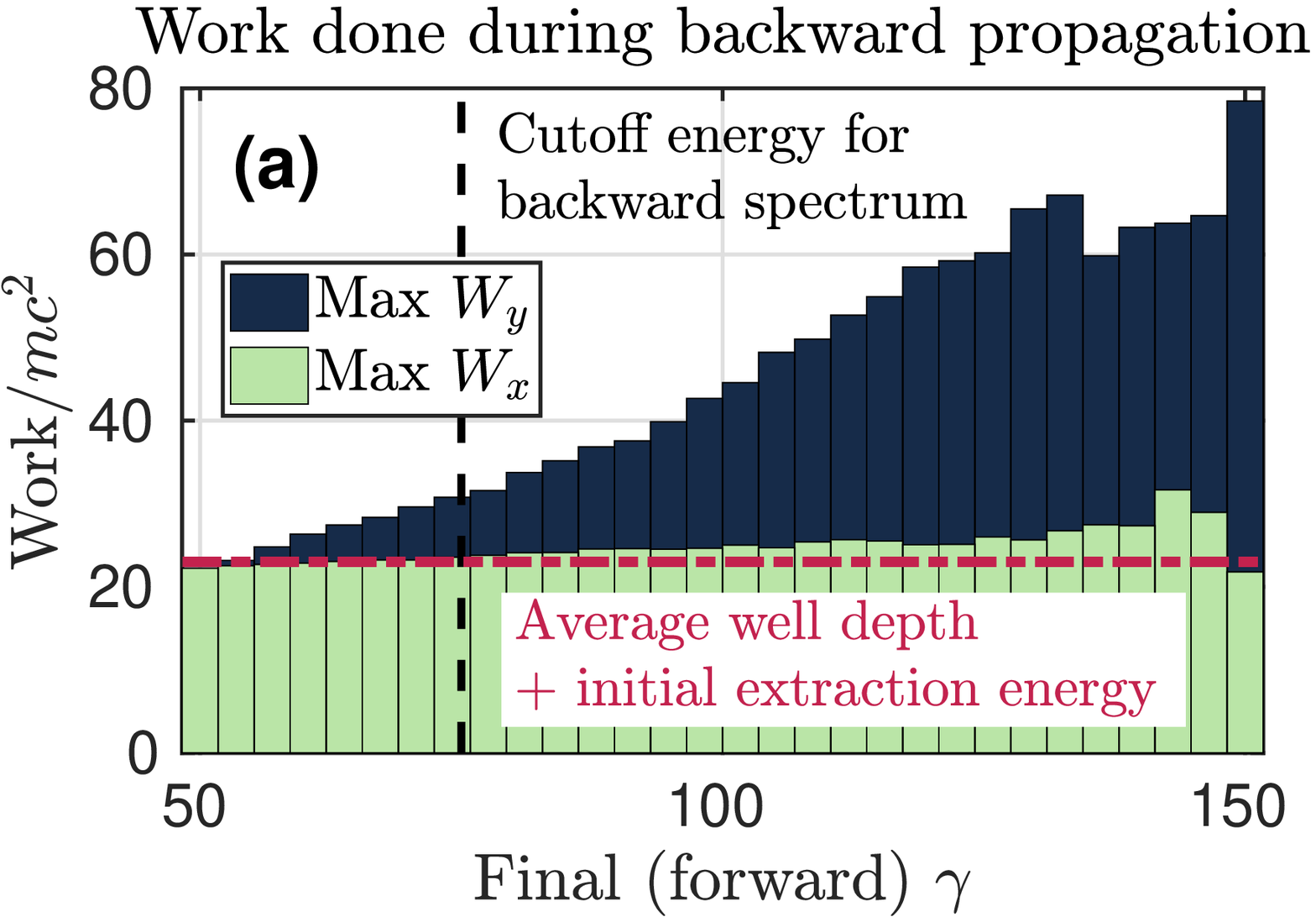}
    \subcaption{}\label{fig:back-work}
    \end{subfigure}
    \begin{subfigure}{0.47\textwidth} 
    \includegraphics[width=0.95\linewidth]{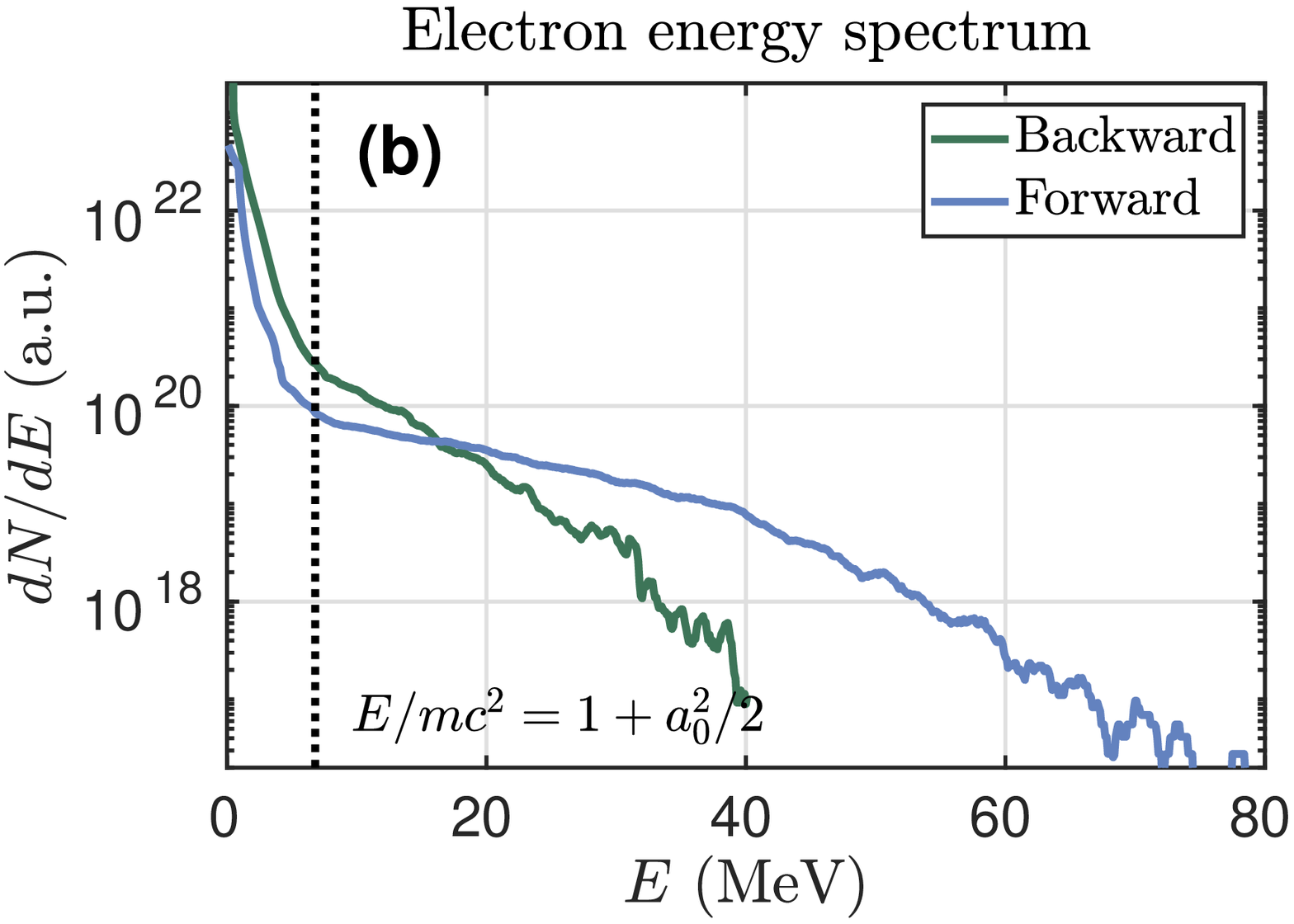}
    \subcaption{}\label{fig:bf-spectrum}
    \end{subfigure}
    \vspace{-1.75\baselineskip}
    \caption{
    Work done on backward-moving electrons and comparison of backward- and forward-moving energy spectra.
    (a) Maximum work done during backward propagation by the longitudinal ($x$, plasma) and transverse ($y$, laser) electric field components, binned by the electron's final (forward-moving) energy.
    (b) Comparison of backward- and forward-moving electron energy spectra, evaluated at the end of the simulation.
    }
    \label{fig:bf-spec-traj}
\end{figure*}

During Step 1 (backward propagation), electrons gain energy from the longitudinal electric field of the plasma and the transverse electric fields of the counter-propagating incident and reflected components of the laser pulse. 
We probe the correlation between acceleration during backward propagation and the electron's final (forward-propagating) energy by binning electrons by their final energy and averaging the maximum backward $W_x$ and $W_y$ of electrons in each final energy bin.
The result is shown in Figure \ref{fig:back-work}. 
No correlation is seen between the electron's final energy and the maximum work done by the longitudinal electric field of the plasma during backward propagation.
The maximum backward $W_x$ is approximately equal to the energy with which the electron is initially extracted from the wall surface plus the depth of the time-averaged electrostatic potential well. 
In contrast, higher final (forward-moving) energy is on average associated with higher backward DLA energy gain ($W_y$). 
We further observe that the majority of the backward DLA energy gain occurs while the longitudinal electric field is decelerating for electrons.

\begin{figure*}
    \centering
    \begin{subfigure}{0.47\textwidth} 
    \includegraphics[width=0.95\linewidth]{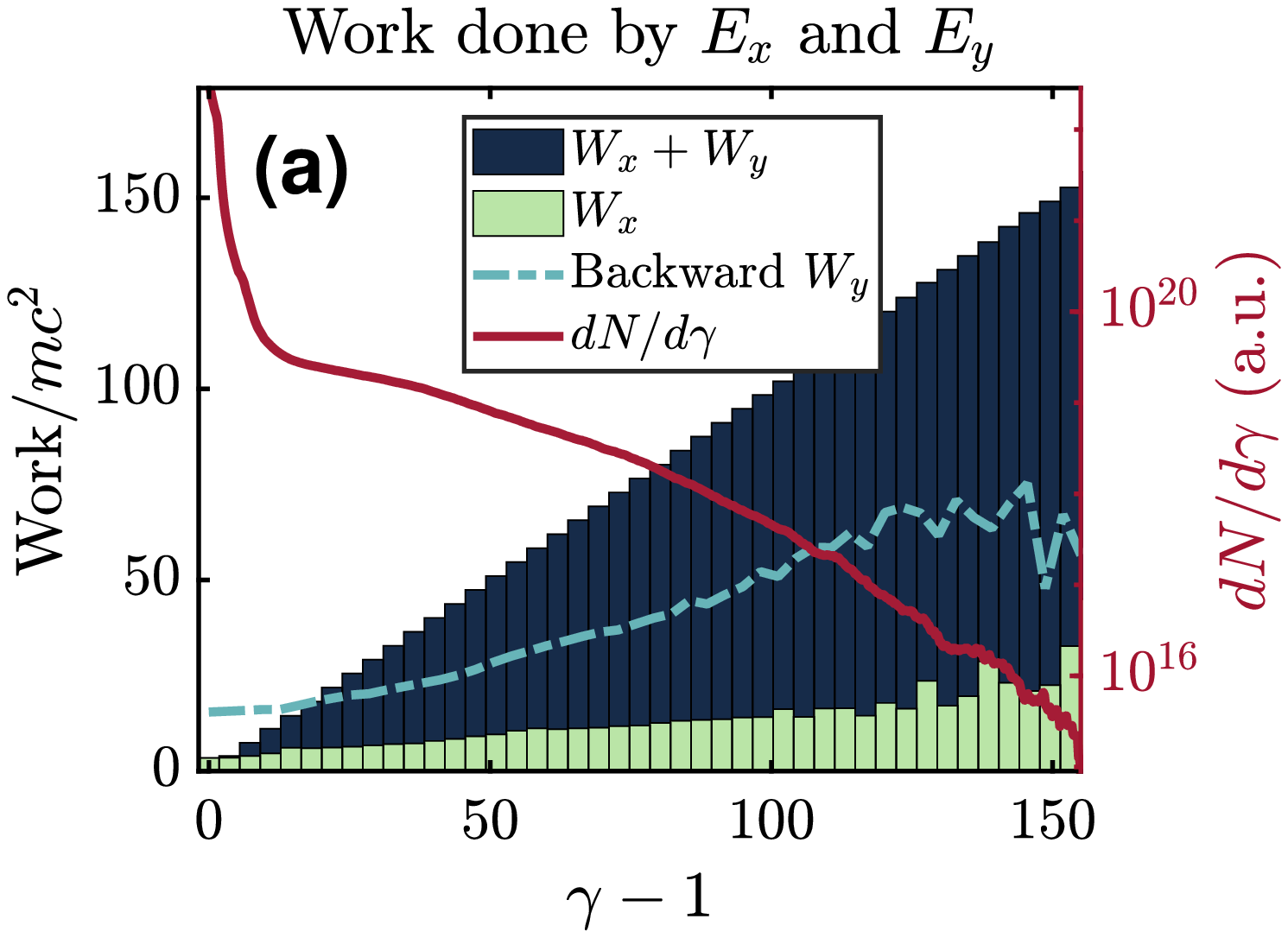}
    \subcaption{} \label{fig:work-bars}
    \end{subfigure}    
    \begin{subfigure}{0.47\textwidth} 
    \includegraphics[width=0.95\linewidth]{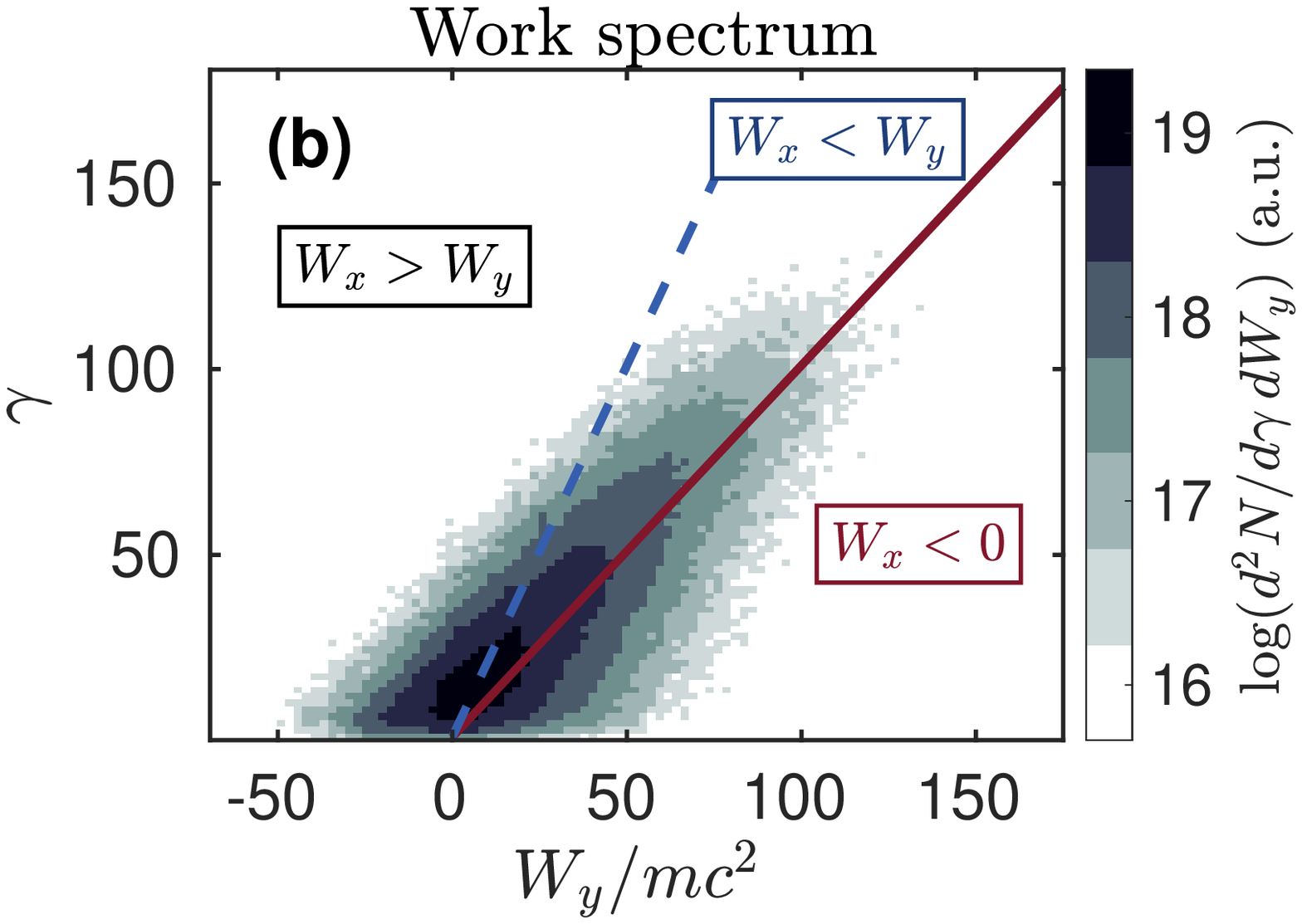}
    \subcaption{} \label{fig:work-spec}
    \end{subfigure}
    \vspace{-1.75\baselineskip}
    \caption{
    Work done on electrons by the the longitudinal ($x$, plasma) and transverse ($y$, laser) electric field components, evaluated at the final time the electron is in the shelf region. (a) Electron energy spectrum and average work, binned by energy, with maximum $W_y$ from backward propagation (Figure \ref{fig:back-work}). 
    (b) Spectral distribution of the work done by the laser field. 
    }
    \label{fig:total-work}
\end{figure*}

Like the backward acceleration stage, the forward electron acceleration of Step 3 is also strongly DLA-dominated. 
A net energy accounting for the electrons (Figure \ref{fig:work-bars}) reveals the longitudinal electric field contributes on average less than 20\% of the final electron energy and is substantially smaller than the energy contribution from backward DLA. 
The longitudinal electric field does net negative work for a substantial fraction of the high energy electrons (Figure \ref{fig:work-spec}).

\begin{figure}
    \centering
    \includegraphics[width=0.95\linewidth]{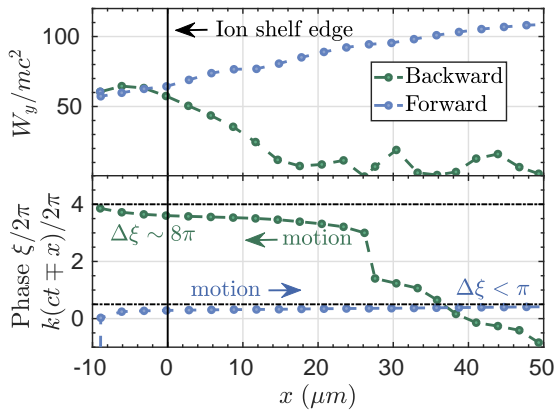}
    \caption{Representative high energy electron trajectory from PIC simulation with (top) the work done on the electron by the laser electric field and (bottom) the phase slip the electron experiences in the co-propagating laser.}
    \label{fig:sample-traj}
\end{figure}

There is a notable difference in the electron spectra (Figure \ref{fig:bf-spectrum}) and the  character of the acceleration during backward and forward propagation. 
We observe a significant difference in the ability of electrons to maintain phase in the co-propagating component of the laser pulse, exemplified by the characteristic high energy electron trajectory shown in Figure \ref{fig:sample-traj}. 
During the forward propagation stage, the phase electrons experience in the incident (forward-propagating) laser pulse changes by less than one period over the entire propagation over the shelf, whereas during the backward propagation stage, the electrons slip in phase by at least several periods with respect to the reflected (backward-propagating) part of the pulse (Figure \ref{fig:sample-traj}, bottom).
The work done on electrons by the laser electric field (Figure \ref{fig:sample-traj}, top) reflects this phase slip, with $W_y$ changing smoothly during forward propagation, but exhibiting cycles of rapid energy gain and loss during backward propagation.

To summarize, we find electron acceleration proceeds in two stages corresponding to forward and backward propagation. Although the character of the stages is different, both stages appear to correspond to DLA alone, without a substantial synergistic contribution from the longitudinal electric field.
As will be discussed in more detail in Sections \ref{sec:DLA-model} and \ref{sec:discuss}, the character of the forward acceleration and the non-involvement of the longitudinal electric field are in contrast with previous explanations for energy gain in multipicosecond laser-solid interaction \cite{sheng2004stochastic_pic, kemp2009psshelf_pond, paradkar2011scalelength, paradkar2012wellheating_theory, wu2016stoch_es,kojima2019electromagnetic}. 
In Section \ref{sec:DLA-model}, we will demonstrate that DLA alone is capable of reproducing the observed acceleration characteristics, including the difference in character between the backward and forward energy gain.

\section{Electron acceleration in counter-propagating laser pulses} \label{sec:DLA-model}

We found in Section \ref{sec:sim-results} that electron acceleration proceeds in two distinct DLA stages, 
neither of which appear to rely on a net energy contribution from the longitudinal electric field.
In this Section, we will explain these observations using a model for direct laser acceleration in counter-propagating laser pulses.

The equations of motion of an electron in counter-propagating linearly $y$-polarized laser pulses in vacuum are (in 1D)
\begin{align} \label{eqn:emotion}
    \begin{split}
        \dfrac{d x}{dt} & = \dfrac{p_x}{\gamma m} \\
        \dfrac{d p_x}{dt} & = -|e| E_x - \dfrac{|e|p_y}{\gamma mc} B_z \\
        \dfrac{d p_y}{dt} & = -|e| E_y + \dfrac{|e|p_x}{\gamma mc} B_z.
    \end{split}
\end{align}
where the choice of $y$-polarization has specified $E_z = B_y = 0$, and we have taken $B_x = 0$ and $p_z = 0$ for consistency with our PIC simulation setup. We then separate $E_y$ and $B_z$ associated with the laser field into separate $+x$- and $-x$-directed components, denoted with subscripts $1$ and $2$, respectively.

We relate $B_{z,1,2}$ to $E_{y,1,2}$ via Maxwell's equations by allowing the laser propagation to correspond to vacuum (i.e. setting the phase velocity equal to the speed of light, which is appropriate based on the phase velocity we observe in the PIC simulation, see Appendix \ref{append:vphi}), and express them in terms of the normalized vector potential $\bm{a}_{1,2}$, with $\bm{a}_{1,2} = a_{1,2} \hat{y}$. We take $a_1$ and $a_2$ to be functions of $t-x/c$ and $t+x/c$ for the $+x$- and $-x$-directed pulses, respectively, and define $a_{1,2}$ by
\begin{align} \label{eqn:vec-a}
\begin{split}
    E_{y,1,2} & = -\dfrac{mc}{|e|} \dfrac{\partial}{\partial t} a_{1,2}\\
    B_{z,1,2} &= \dfrac{mc}{|e|} \dfrac{\partial}{\partial x} a_{1,2}.
\end{split}
\end{align}
Substituting these expressions in Equations \ref{eqn:emotion}, we readily obtain an integral of the motion (as per Ref. \onlinecite{kruger1976dla}) 
\begin{equation} \label{eqn:py-int}
    \dfrac{d}{dt} \left(\dfrac{p_y}{mc} - a_1 - a_2 \right) = 0.
\end{equation}

We take $E_x = 0$ for the purposes of our analysis. 
An accelerating longitudinal electric field dynamically reduces the electron dephasing rate in a single laser pulse \cite{robinson2013dephasing} and it has been suggested that $E_x$ could play an important role in electron acceleration by picosecond pulses \cite{paradkar2011scalelength}. However, we find no evidence for a substantive synergistic energy gain from $E_x$ during either backward propagation (see Section \ref{sec:sim-results}). In this model, we treat $E_x$ as contributing to the initial longitudinal momentum to determine what effect, if any, it can have on the electron energy gain.

Combining Equations \ref{eqn:emotion}, \ref{eqn:vec-a}, and \ref{eqn:py-int}, the equations of motion for an electron in this system are
\begin{align} \label{eqn:DLA-model}
    \begin{split}
        \dfrac{d x}{dt} & = \dfrac{p_x}{\gamma m} \\
        \dfrac{d p_x}{dt} & = \dfrac{p_y}{\gamma} \left(\dfrac{da_1}{d\left(t-x/c\right)} - \dfrac{da_2}{d\left(t+x/c\right)}\right) \\
        p_y & = p_{y,0} + a_1 mc + a_2 mc - a_{1,0}mc - a_{2,0}mc \\
        a_{1,2} & = a_0 \sin \left(ck \left[t \mp x/c \right] + \phi_{1,2}\right).
    \end{split}
\end{align}
The electron is injected into the counter-propagating pulses at $x=0$, $t=0$ with a starting phase in each laser pulse given by $\phi_{1,2}$. 
We take the initial longitudinal momentum $p_{x,0} \ge 0$ to probe the importance of pre-acceleration. 
We also allow for an initial $p_{y,0}$ to represent the modulation of the transverse momentum by the counter-propagating laser pulses during pre-acceleration. 
$p_{x,0} \ge 0$ is also suitable to probe backward propagation due to the symmetry of Equations \ref{eqn:DLA-model} as $(x,\bm{p}) \to (-x,-\bm{p})$ (with suitable modification of the initial laser phases).
In the backward case, the range of initial $p_{x,0}$ and $p_{y,0}$ represents the initial extraction of the electron from the surface and acceleration by the electrostatic potential.
We perform scans over $\phi_{1,2}$, $p_{x,0}$ and $p_{y,0}$ to explore the range of initial conditions electrons may experience in the PIC simulation.

\begin{figure*}
    \centering
    \begin{subfigure}{0.32\textwidth} 
    \includegraphics[width=0.95\linewidth]{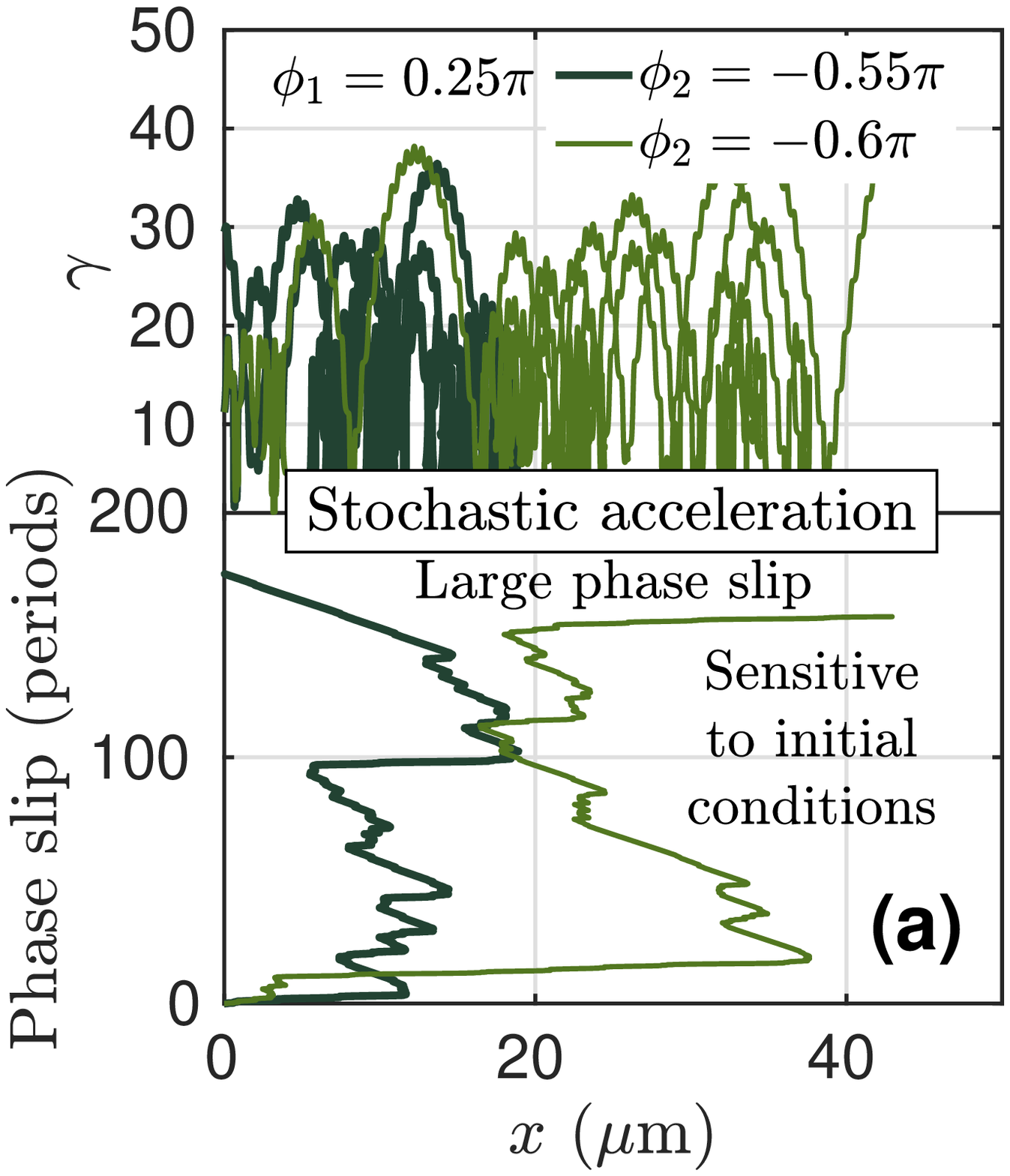}
    \subcaption{} \label{fig:traj-2-stoc}
    \end{subfigure}    
    \begin{subfigure}{0.32\textwidth} 
    \includegraphics[width=0.95\linewidth]{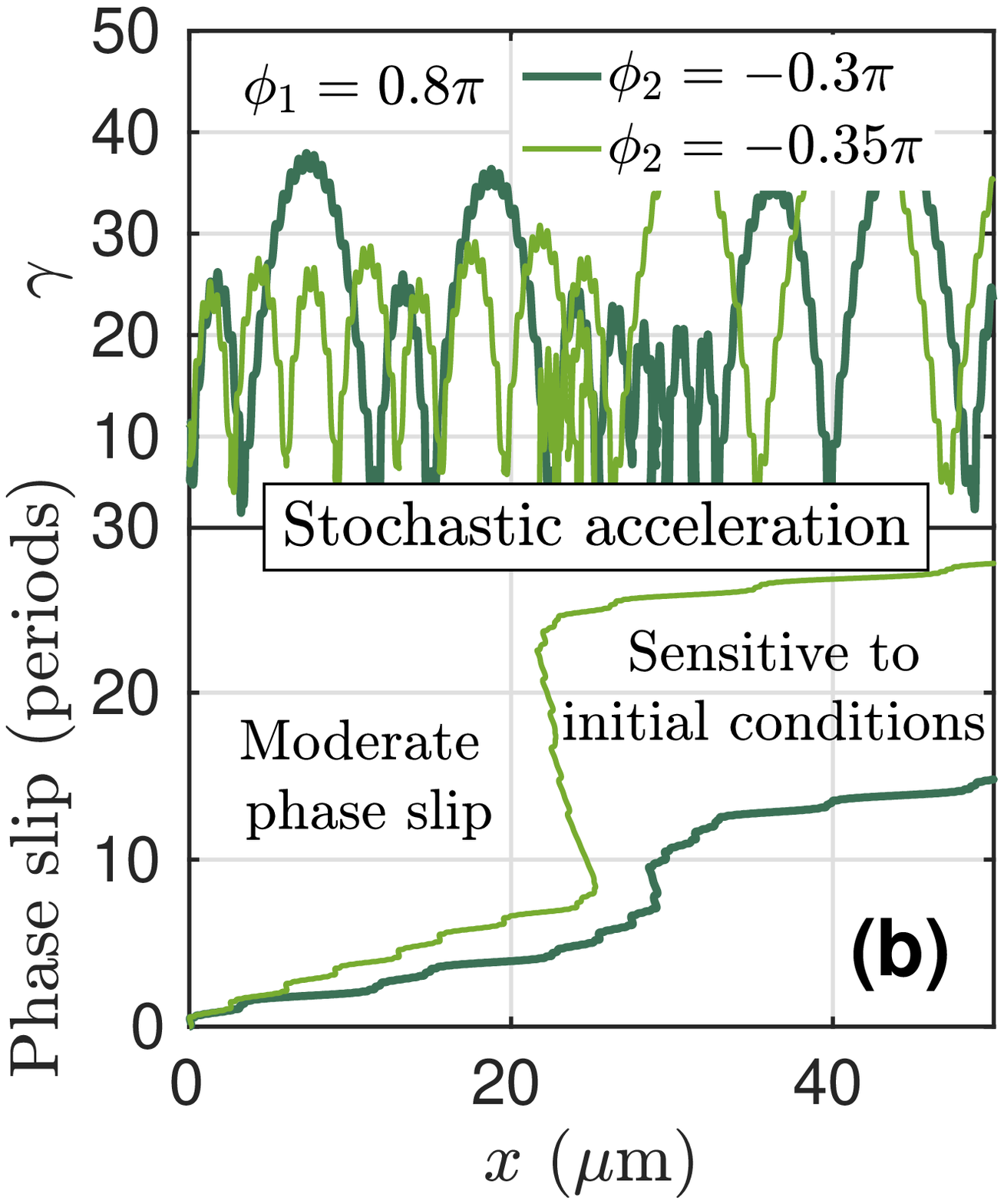}
    \subcaption{} \label{fig:traj-2-mid}
    \end{subfigure}
    \begin{subfigure}{0.32\textwidth} 
    \includegraphics[width=0.95\linewidth]{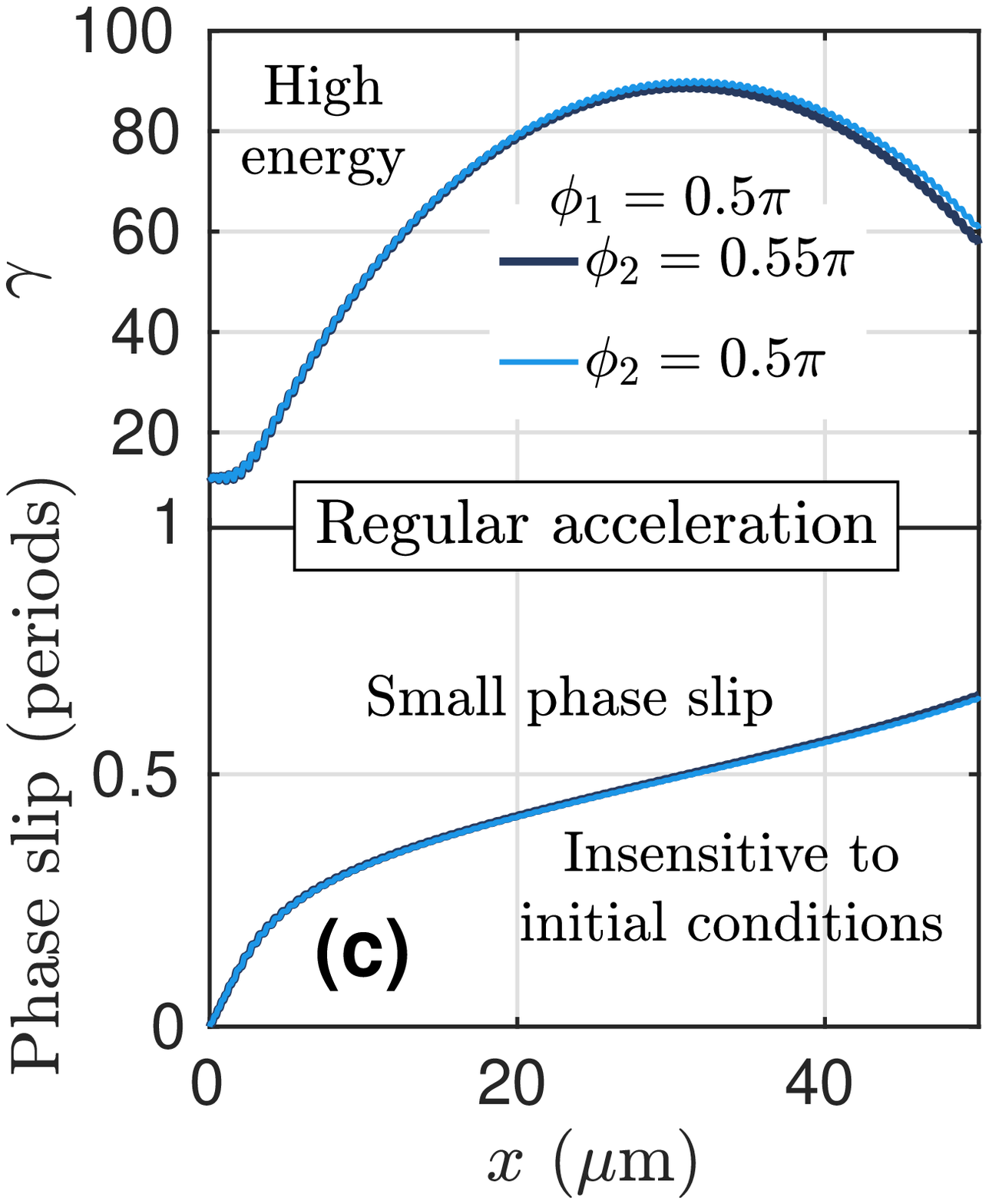}
    \subcaption{} \label{fig:traj-2-reg}
    \end{subfigure}    
    \vspace{-1.75\baselineskip}
    \caption{
    Example electron trajectories for initial longitudinal momentum $p_{x,0}=10$ and transverse momentum $p_{y,0}=5$. The two trajectories in each figure were obtained by fixing the initial phase in one pulse ($\phi_1$) and varying the other ($\phi_2$) by $0.05 \pi$. (a) Stochastic acceleration consistent with stochastic heating (large phase slip). (b) Stochastic acceleration with moderate phase slip. (c) Regular acceleration.
    }
    \label{fig:traj-2-laser}
\end{figure*}

The electron position and momentum are evolved according to Equations \ref{eqn:DLA-model} for a time of 200 laser periods using MATLAB's ode113 solver with relative error tolerance of $10^{-13}$. 
We find that electron acceleration is accomplished in two distinct regimes.
Certain initial conditions lead to stochastic motion of the electron (e.g. Figures \ref{fig:traj-2-stoc} and \ref{fig:traj-2-mid}), while others produce motion which appears regular (e.g. Figure \ref{fig:traj-2-reg}).
Stochastic electron trajectories are difficult to compute exactly \cite{robinson2019solvers} 
and it was not possible to achieve convergence for some of the initial conditions leading to stochastic motion. 
The intent of modeling stochastic motion in this work is not to present exact trajectories, but to obtain characteristic values for the maximum $\gamma$-factor of electrons and the amount the electron slips in phase in each laser pulse over the shelf length. 
For the purposes of the subsequent analysis, we have verified that the solver preserves key physical features of electron motion in counter-propagating pulses.
Over our range of initial conditions, the solver is converged for the regular trajectories, exhibits the phase space filling behavior needed to capture the maximum $\gamma$-factor and the rapid divergence of neighboring trajectories in the stochastic regime, 
and preserves the phase-space separation between stochastic and regular motion we expect on the basis of Poincar\'e surface of section plots (as shown, e.g. in Ref. \onlinecite{mendonca1983stochastic}).

We distinguish the stochastic regime from the regular regime in this work on the basis of the motion of the electron, not as $t \to \infty$, as would formally define chaotic/stochastic behavior, but over a propagation distance corresponding to the shelf length. 
The whole of the electron bounce trajectory in the PIC simulation (both backward and forward motion) takes on average 500 fs (150 laser cycles). In this section, we model electron acceleration in counter-propagating systems for up to 667 fs (200 laser cycles), which is sufficient for electrons to be displaced by more than the shelf length for nearly all the initial conditions we evaluate. 
In contrast, past treatments of stochastic heating have typically been based on conditions where electrons interact with the counter-propagating electromagnetic waves over a relatively long time to allow electrons to explore the full range of accessible phase space (corresponding to hundreds to thousands of laser periods worth of phase slip, for example References \onlinecite{sheng2004stochastic_pic,kemp2009psshelf_pond,bochkarev2019stoch}).

In our model, electrons slip anywhere from more than one hundred to less than one period in the co-propagating laser over the shelf length.
We find that the hallmarks of stochasticity (e.g. the divergence of neighboring orbits signified by the sensitive dependence on initial conditions shown in Figures \ref{fig:traj-2-stoc} and \ref{fig:traj-2-mid}) are associated with electrons which slip by tens of more of periods in phase.
The usual stochastic heating picture is clearly applicable to electrons which slip by hundreds of laser periods in phase (e.g. for trajectories in Figure \ref{fig:traj-2-stoc}), but may not be for electrons which only slip tens of periods in phase slip (e.g. Figure \ref{fig:traj-2-mid}). We term both of these cases "stochastic," and contrast them with "regular" electrons.
In the regular regime, trajectories do not depend sensitively on the initial conditions, electrons slip in phase by at most a few laser periods over the shelf length, and the peak energy can be higher (e.g. Figure \ref{fig:traj-2-reg}).

\begin{figure*}
    \centering
    \begin{subfigure}{0.47\textwidth}
    \includegraphics[width=0.95\linewidth]{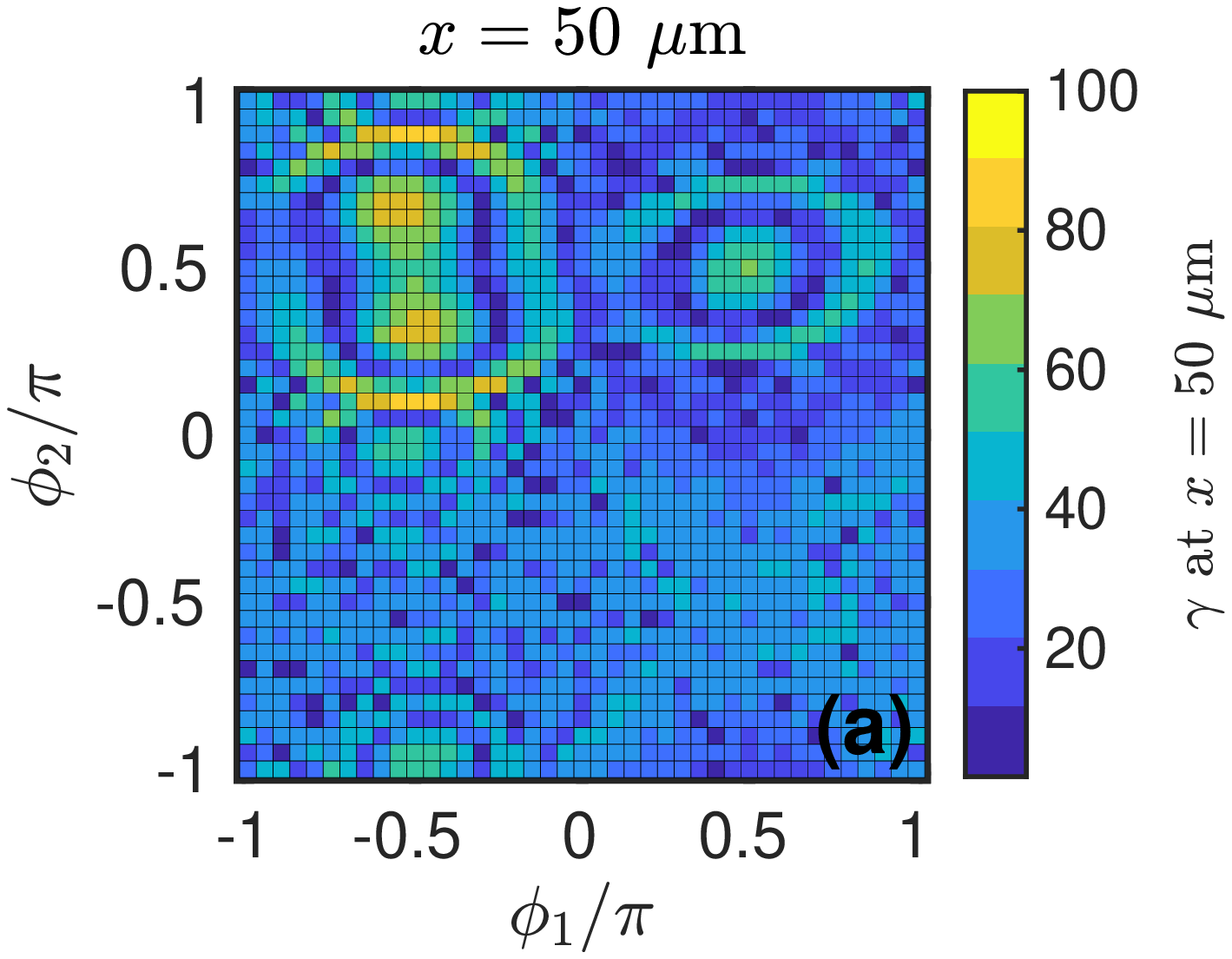}
    \subcaption{} \label{fig:plane-g-phase}
    \end{subfigure}    
    \begin{subfigure}{0.47\textwidth}
    \includegraphics[width=0.95\linewidth]{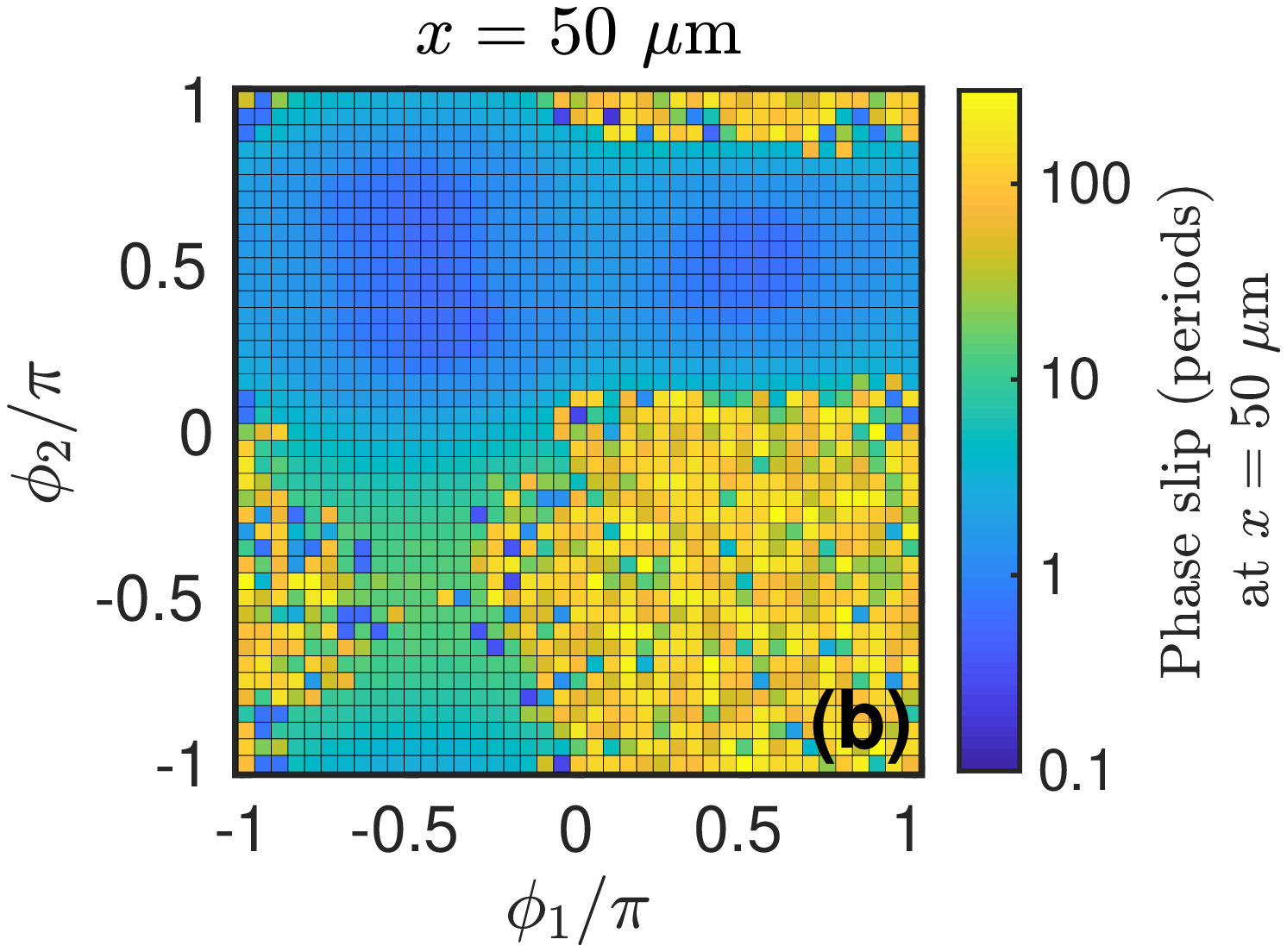}
    \subcaption{} \label{fig:plane-ph-phase}
    \end{subfigure}
    \vspace{-1.75\baselineskip}
    \caption{
    Evaluation of electron trajectories in counter-propagating lasers after propagation over the shelf length ($x=50$ $\mu$m), scanning over the initial laser phases $\phi_1$ and $\phi_2$, with initial longitudinal momentum $p_{x,0}/mc=10$ and transverse momentum $p_{y,0}/mc=5$. (a) $\gamma$-factor after propagation over the shelf length and (b) corresponding phase slip of the electron in the $+x$-propagating laser.
    }
    \label{fig:plane-phase}
\end{figure*}

The distinct characteristics of the stochastic and regular regimes are easy to identify in a scan over the initial phases $\phi_1$ and $\phi_2$.
Figure \ref{fig:plane-g-phase} shows the $\gamma$-factor as the electron crosses $x = 50$ $\mu$m for the initial momentum conditions $p_{x,0}/mc = 10$, $p_{y,0}/mc = 5$.
We see that for $\phi_1 \gtrsim 0$ and $\phi_2 \lesssim 0$, both $\gamma$ and the phase of the electron in the co-propagating laser (shown in Figure \ref{fig:plane-ph-phase}) vary from pixel to pixel in our calculation (one pixel corresponds to $\Delta \phi/\pi = 0.05$), revealing the sensitive dependence on initial conditions characteristic of the stochastic regime. 
By contrast, the $\gamma$-factor and the phase slip vary smoothly in the regular regime (the remainder of the phase scan).
We find that the highest energy electrons correspond to the regular regime.
These electrons have begun to decelerate by the time they reach $x=50$ $\mu$m, which suggests that additional pre-acceleration can improve the electron energy.

We repeat the scan over the starting $\phi_{1,2}$ for different values of $p_{x,0}$ and $p_{y,0}$ and find that the regular regime is made accessible through sufficient pre-acceleration, though regularity also requires the electron to have favorable initial phases in the two pulses (as can be seen, for example, in Figure \ref{fig:plane-ph-phase}).
When the electron has initial forward momentum $p_{x,0}/mc \lesssim a_0 = 5$, the motion is stochastic for all initial choices of $\phi_{1,2}$, while for $p_{x,0}/mc \gtrsim a_0$, as shown in Figures \ref{fig:traj-2-laser} and  \ref{fig:plane-ph-phase}, certain initial choices in $\phi_{1,2}$ produce stochastic behavior, while others result in regular acceleration. The range of initial conditions producing regular acceleration increases with increasing $p_{x,0}$.

\begin{figure}
    \centering
    \begin{subfigure}{0.5\textwidth}
    \includegraphics[width=0.95\linewidth]{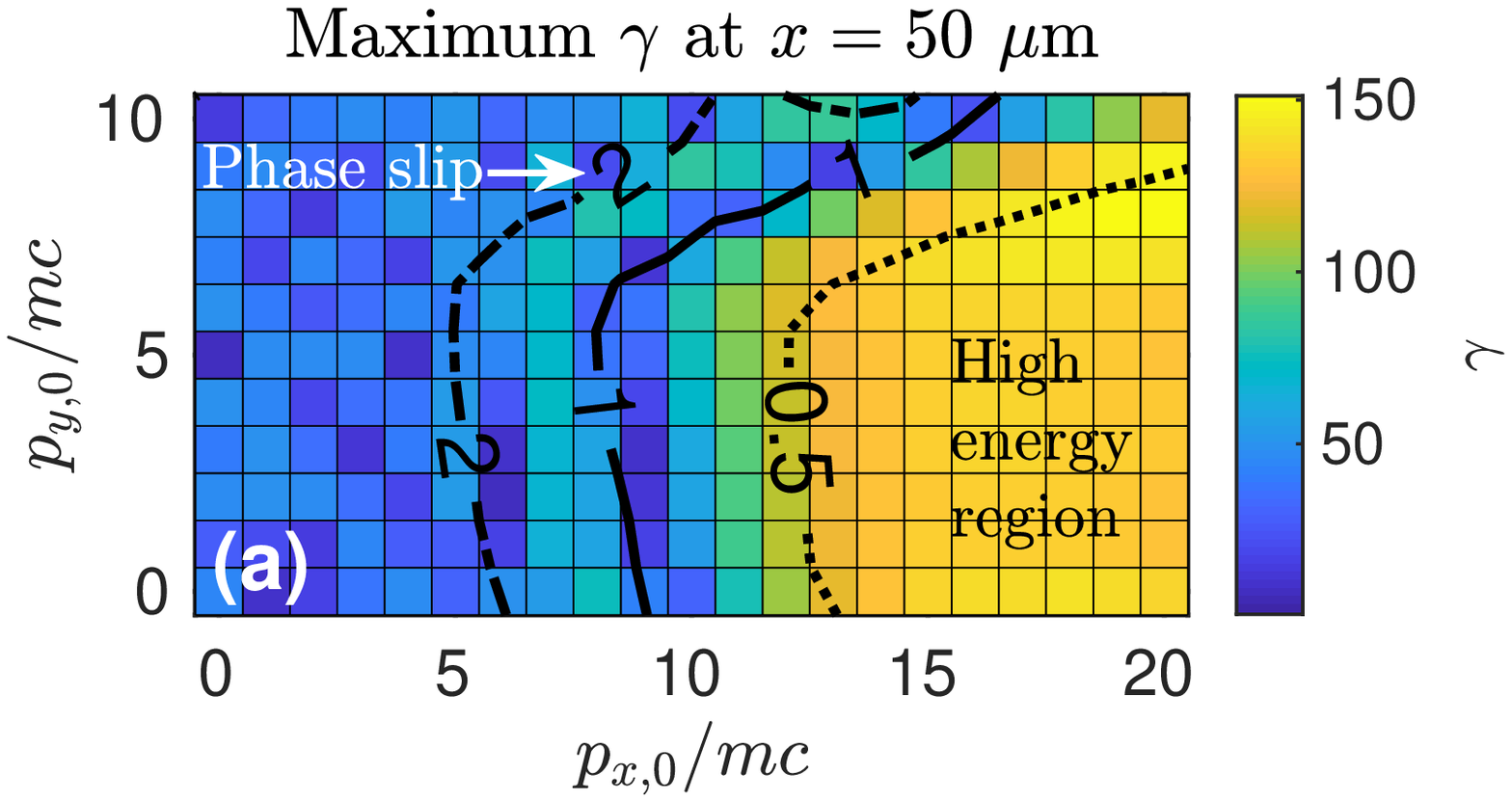}    
    \subcaption{}\label{fig:plane-g-pxpy}
    \end{subfigure}

    \vspace{-1.5\baselineskip}
    \begin{subfigure}{0.5\textwidth} 
    \includegraphics[width=0.95\linewidth]{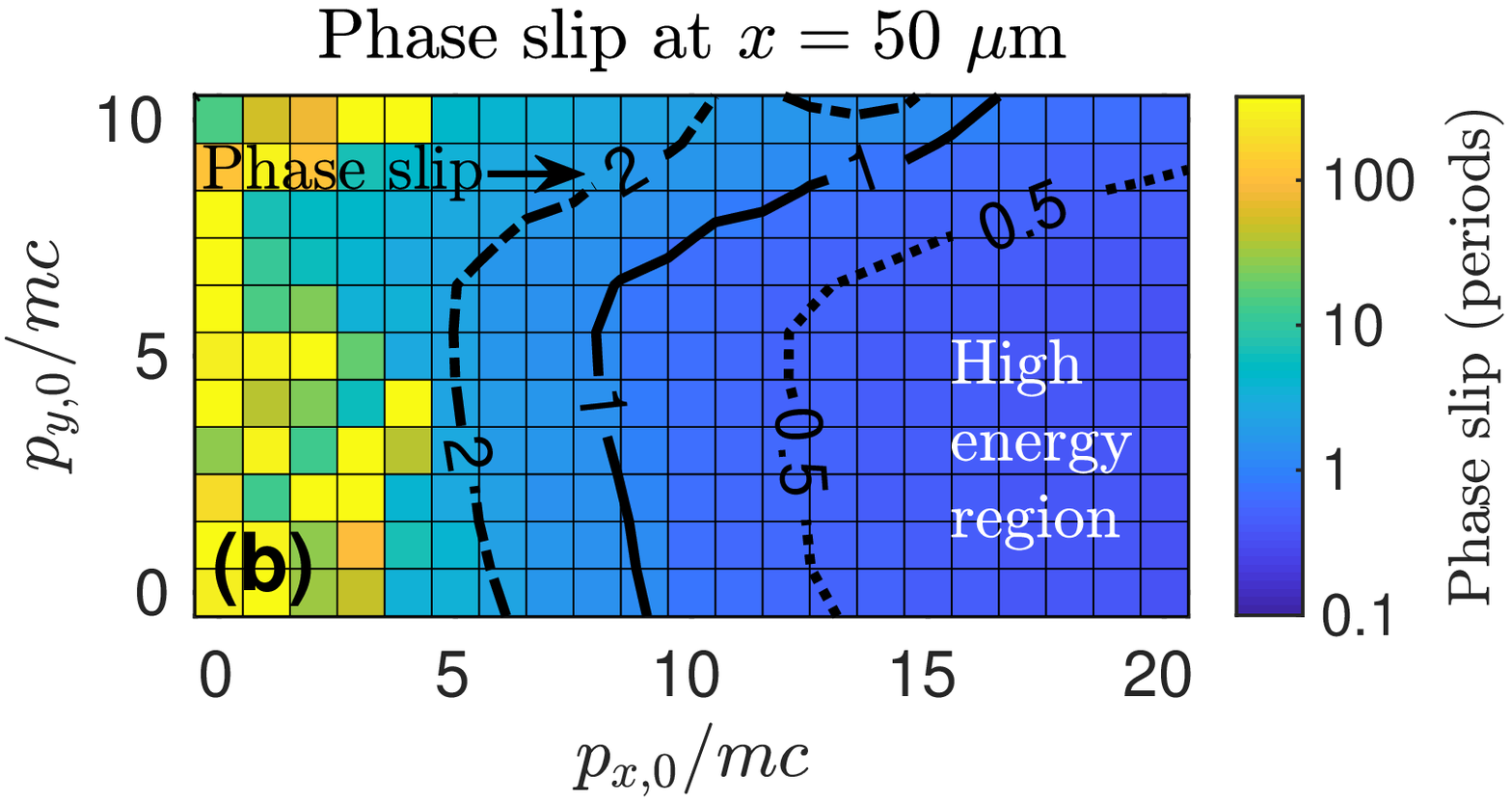}    
    \subcaption{}\label{fig:plane-ph-pxpy}
    \end{subfigure}

    \vspace{-1.75\baselineskip}
    \caption{
    Highest energy electron trajectories in counter-propagating lasers evaluated after after propagation over the shelf length ($x=50$ $\mu$m) for a range of initial longitudinal ($p_{x,0}$) and transverse ($p_{y,0}$) momenta. 
    (a) Maximum achievable $\gamma$-factor at $x=50$ $\mu$m and (b) corresponding phase slip in the $+x$-propagating laser for different starting momenta. Black contours on both (a) and (b) indicate the phase slip in periods.
    } \label{fig:plane-pxpy}
\end{figure}

We also find that sufficient pre-acceleration enables the regular regime to produce higher electron energy at $x=50$ $\mu$m than the stochastic regime. 
Figures \ref{fig:plane-g-pxpy} and \ref{fig:plane-ph-pxpy} show the maximum $\gamma$-factor and associated phase slip in the forward pulse obtained after propagation over the shelf length, compiled by evaluating over 40,000 electron trajectories ($\Delta \phi = 0.1$ for the scan over $\phi_{1,2}$). 
The maximum achievable $\gamma$-factor in this system corresponds to regular acceleration with small phase slip ($\Delta \xi/2 \pi \lesssim 0.5$) and requires $p_{x,0}/mc \gtrsim 15$ for our range of initial $p_{y,0}$.
We also see that pre-acceleration only systematically reduces phase slip and increases the maximum electron energy in the regular regime, and does not in the stochastic regime.

We now compare qualitatively the stochastic and regular acceleration predicted by our counter-propagating DLA model to the PIC simulation results.
The backward acceleration stage exhibits features of stochastic acceleration, including rapid cycles of energy gain and loss with a corresponding multi-period phase slip.
Furthermore, stochasticity would explain the lack of correlation between the backward DLA energy gain and the backward pre-acceleration provided by the longitudinal electric field. 
The forward acceleration stage, on the other hand, agrees well with regular acceleration.
High energy electrons exhibit minimal phase slip in the forward-propagating laser over the shelf length and higher final energy is associated with higher pre-acceleration by backward DLA.

The electron energy gain process we observe in the simulation can be summarized as follows. Electrons are extracted from the solid density plasma surface and backward accelerated by the electrostatic potential well and by stochastic DLA. This is not a synergistic process and DLA is the dominant contributor to the energy gain. Next, the electron bounces in the electrostatic potential and the backward energy gain is transformed into a forward pre-acceleration. The backward DLA energy gain can be sufficient for the electron to then be forward accelerated by non-stochastic DLA. When the forward acceleration is non-stochastic, electrons can gain higher energy than is produced by stochastic heating. In this single-bounce process, the only important role of the electrostatic potential is to reflect electrons during the bounce. We find the key to energetic electron production is laser reflection from the critical density surface.

\section{Summary and Discussion} \label{sec:discuss}

We investigated the production of high energy electrons in a system representative of the plasma and laser conditions produced by multipicosecond laser-solid interaction. 
In particle-in-cell simulations designed to isolate the electron dynamics occurring near the peak of a multipicosecond laser pulse, we observed high energy electron production via a two-stage acceleration process involving an initial backward phase and a final forward phase, which we demonstrated are consistent with stochastic and non-stochastic direct laser acceleration, respectively. The only required role of the electrostatic potential in this process is to reflect high-energy electrons between the backward and forward stages.

Previous work on electron heating by multipicosecond laser pulses incident on solid density targets has focused on stochastic heating \cite{sheng2004stochastic_pic,kemp2009psshelf_pond}, requires multiple bounces in the electrostatic potential \cite{paradkar2012wellheating_theory}, or involves a substantive energy contribution from the electrostatic potential \cite{paradkar2011scalelength,wu2016stoch_es,kojima2019electromagnetic}. We have demonstrated electron acceleration can be achieved in a two-stage, \textit{single}-bounce DLA process in which \textit{only one} stage is stochastic and the longitudinal electric field is \textit{not required} to explain the observed energy gain. 
The conclusion we draw is that, for our model conditions, the key factor driving the acceleration of electrons to high energy is reflection of the laser pulse from the dense plasma surface.

\begin{figure*}
    \centering
    \begin{subfigure}{0.47\textwidth}
    \includegraphics[width=0.95\linewidth]{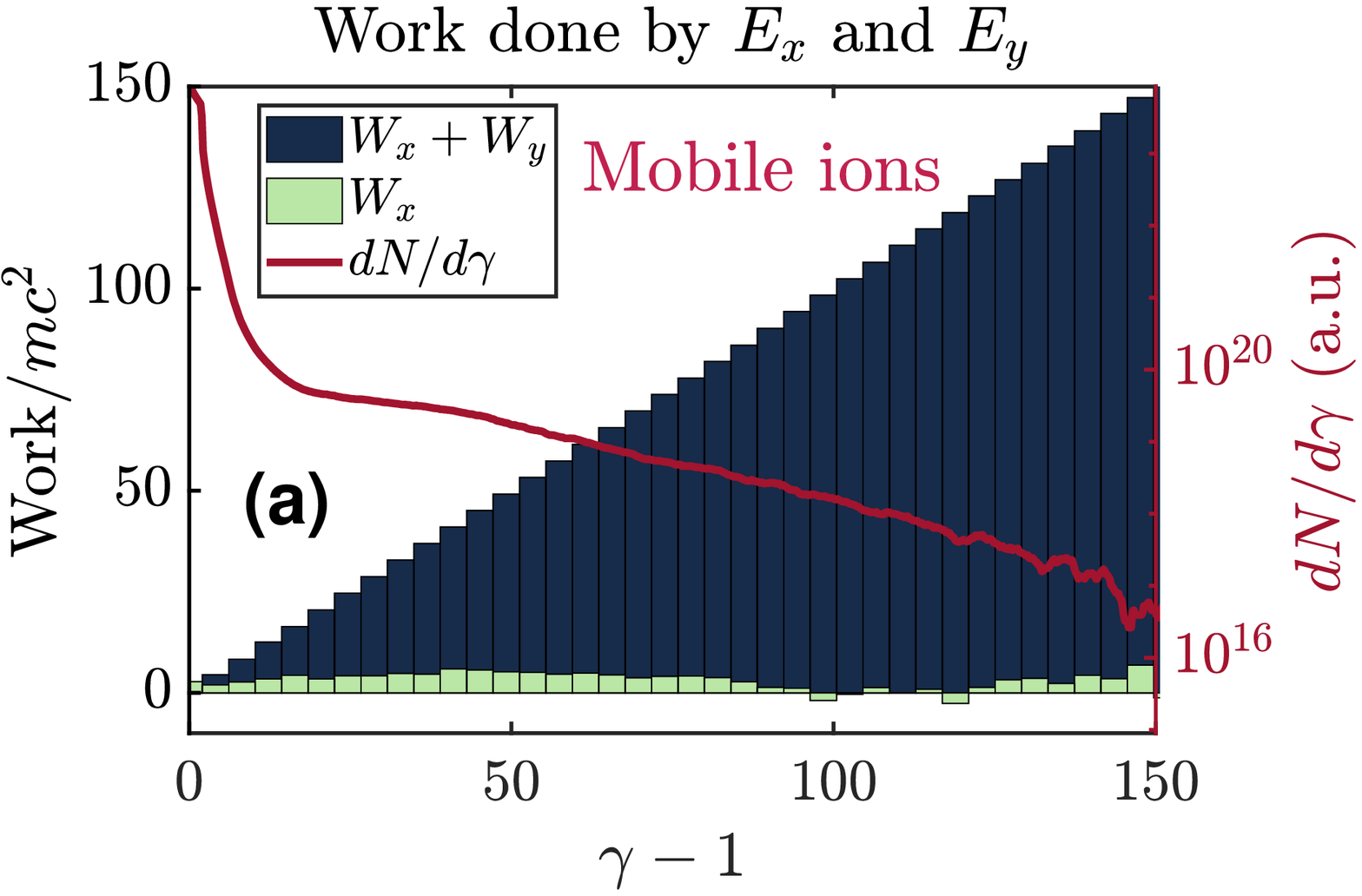}
    \subcaption{} \label{fig:work-bars-mobile}
    \end{subfigure}    
    \begin{subfigure}{0.47\textwidth}
    \includegraphics[width=0.95\linewidth]{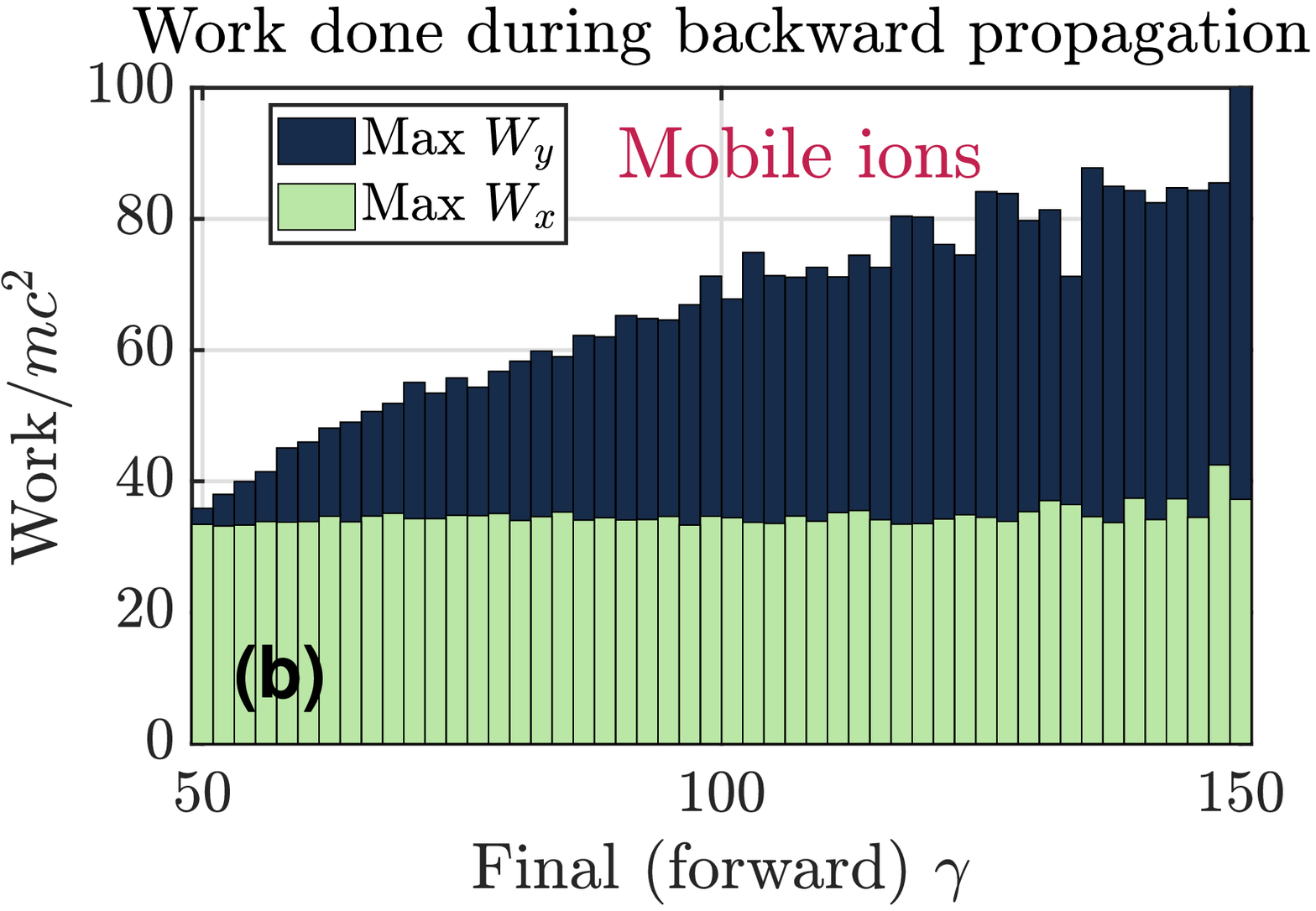}
    \subcaption{} \label{fig:work-back-mobile}
    \end{subfigure}
    \vspace{-1.75\baselineskip}
    \caption{
    Work done on electrons by the the longitudinal ($x$, plasma) and transverse ($y$, laser) electric field components, evaluated at the final time the electron is in the shelf region for the case with mobile ions.
    (a) Electron energy spectrum and average work, binned by energy. Compare to Figure \ref{fig:work-bars}.
    (b) Maximum work done during backward propagation, binned by final energy. Compare to Figure \ref{fig:back-work}
    }
    \label{fig:work-mobile}
\end{figure*}

The intent of this work is to provide a framework for future analysis. To this end, we have neglected several potentially important aspects of the physics of picosecond laser-solid interaction which should be included in future models to evaluate the applicability of the two-stage acceleration process we describe in more complex situations. While performing an in-depth analysis is beyond the scope of this work, we now discuss briefly the robustness of the acceleration process we describe to some of these additional considerations.

First, we expect the backward pre-acceleration required to enable high energy forward regular acceleration to depend on the laser intensity and the shelf length, both of which will evolve over the course of a multipicosecond laser pulse. Conceptually, we require that the stochastic energy gain be sufficient to enable the forward acceleration to access the regular regime and that the regular regime be able to deliver higher electron energy than the stochastic regime over an acceleration distance corresponding to the shelf length. We demonstrate the feasibility of this process for an additional set of conditions in Appendix \ref{append:other-parameters}. We expect our acceleration mechanism to be conceptually valid over some portion of the parameter space, within which it may be possible to optimize the preplasma for electron heating. Experimentally, one could consider using one laser pulse to generate the preplasma and begin the shelf formation and another to perform electron acceleration, akin to the approach used in Refs. \onlinecite{peebles2017preplasma,peebles2018preplasma}.

Second, we note that 1D geometry and ion immobility can artificially suppress laser absorption at the wall surface \cite{kemp2012ps_2d,iwata2018hole_boring}. 
Similar to the analysis of Appendix \ref{append:other-parameters}, we have verified that the electron acceleration process we observe remains feasible even with relatively high absorption, for example, with the $\sim 75\%$ absorption ($50 \%$ reduction in $a_\mathrm{reflected}$) observed in Ref. \onlinecite{kemp2012ps_2d}. The dominance of this energy gain process should be checked in systems with higher absorption.

Third, ion immobility and the temporally semi-infinite profile are most applicable under conditions where the ion shelf evolution does not substantially change the character of electron acceleration. To demonstrate that the two-stage character of electron acceleration is robust to ion mobility, we conduct an additional PIC simulation where we allow the ions to be mobile, fully ionized aluminum. In this simulation, the ion shelf density drops by approximately a factor of 2 over 1 picosecond, but the time-averaged electrostatic potential and the forward-moving electron energy spectrum are not substantially affected. While in-depth analysis of the effects of ion mobility is beyond the scope of this work, we verify that key features of the acceleration are unchanged. High energy electrons are still predominantly DLA-accelerated (Figure \ref{fig:work-bars-mobile}) through two stages of DLA where the final forward stage exhibits non-stochastic character.  Additionally, high final electron energy is still correlated with high backward DLA energy gain, while there is still no correlation between energy gain from the longitudinal electric field and energy gain during either DLA stage (Figure \ref{fig:work-back-mobile}).

We additionally note that 1D geometry is most applicable to conditions under which the preplasma shelf is short compared to its transverse extent, as may be produced by the large focal spot size available from high-power laser facilities including LFEX \cite{Miyanaga2006LFEX}, LMJ-PETAL 
\cite{batani2014PETAL}, NIF ARC \cite{crane2010NIF_ARC}, and OMEGA EP \cite{maywar2008OMEGA_EP}. In 2D and 3D geometry, the laser pulse may generate a plasma channel, introducing an additional timescale associated with the time it takes for an electron to transit the channel transversely. 
For the non-stochastic DLA we observe during forward motion, the electron motion is predominantly forward-directed and we expect the limited transverse size to introduce only secondary effects on the forward DLA process if the spot is sufficiently large. The backward DLA process should be re-evaluated in the context of 2D and 3D simulations.

As a final note, the two-stage DLA process we have described relies on the phase of electrons being either randomized or favorably preserved during the transition from the stochastic backward DLA to the regular forward DLA. However, if the electric field is sufficiently strong, the electrostatic potential appears as a hard wall and the electron phases are not randomized during the bounce, which could in principle result in electrons never being injected into a favorable phase for regular acceleration.  This analysis, as well as an analysis of the conditions under which stochastic heating may occur during the bounce itself, will be the subject of a follow-up publication.

\section{Acknowledgements}

This research was supported in part by the DOE Office of Science under Grant No. DE-SC0018312. 
K. Weichman was supported in part by the DOE Computational Science Graduate Fellowship under Grant No. DE-FG02-97ER25308. 
Particle-in-cell simulations were performed using EPOCH \cite{arber2015epoch},  developed under UK EPSRC Grant Nos. EP/G054940, EP/G055165, and EP/G056803.
This work used HPC resources of the Texas Advanced Computing Center (TACC) at the University of Texas at Austin and the National Energy Research Scientific Computing Center (NERSC), a U.S. Department of Energy Office of Science User Facility operated under Contract No. DE-AC02-05CH11231.
Data collaboration was supported by the SeedMe2 project \cite{chourasia2017seedme} (http://dibbs.seedme.org).

\appendix

\section{Phase velocity}
\label{append:vphi}

In this Appendix, we demonstrate that the phase velocity is sufficiently close to the speed of light that it is appropriate to take $v_\phi = c$ for the purposes of our analysis.

First, we determine the upper limit on the phase velocity we can reasonably neglect. 
It is important for the electron trajectory analysis in Section \ref{sec:sim-results} (Figure \ref{fig:sample-traj}) and for highest energy electrons in the DLA model in Section \ref{sec:DLA-model} (Figure \ref{fig:plane-pxpy}) that we capture accurately whether a high energy electron has transitioned from the accelerating to the decelerating phase in its co-propagating laser pulse, i.e. the maximum error in the phase slip we can tolerate is 1/4 laser period.

The phase of an electron in the $+x$-directed laser pulse incorporating and neglecting the phase velocity are given by
\begin{align}
\begin{split}
    \xi_{v_\phi} & = k\left(v_\phi t - x\right) \\ 
    \xi_c & = k\left(c t - x\right),
\end{split}
\end{align}
respectively.
The difference in phase introduced by neglecting the phase velocity as an electron travels over the shelf distance is therefore
\begin{equation}
    \xi_{v_\phi} - \xi_c = k\left(v_\phi - c\right)t < \pi/2.
\end{equation}
In the PIC simulation, electrons traverse the shelf in less than 500 fs on average, therefore we require $v_\phi/c-1 \lesssim 10^{-3}$.

Next, we estimate the phase velocity of the forward-propagating laser in the PIC simulation, which we find differs substantially from the cold neutral plasma value. 
Established graphical methods for directly measuring the phase velocity in PIC simulation \cite{arefiev2016vphi,willingale2018vphi} are challenging to apply and interpret in the presence of interference patterns, such as results from the strong laser reflection in our simulation.
Instead, we constrain the phase velocity by examining the laser work ($W_y$) done on high energy electrons as they slip in phase. 

The basis of this approach is that an electron injected into an initially accelerating phase in the forward-propagating laser will reach a maximum energy and then begin to decelerate once the electron slips in phase by $1/4$ period. Whether a transition from acceleration to deceleration is visible in $W_y$ for a high energy electron indicates whether the electron slips by more or less than $1/4$ laser period. We can simultaneously calculate the phase slip based on the particle position and time and determine whether it agrees with $W_y$.

For convenience, the following analysis is based on the electron trajectory of Figure \ref{fig:sample-traj}, though other electron trajectories from the simulation may serve to more tightly constrain $v_\phi$. We see based on $W_y$ that this electron does not experience a transition from accelerating to decelerating phase, therefore we consider the actual phase slip to be less than $0.25$ laser period. Simultaneously, we calculate the phase slip over the course of the acceleration (starting from $x=-5$ $\mu$m in Figure \ref{fig:sample-traj}) to be $\sim 0.15$ laser period, occurring over a time of $\sim 55$ laser cycles. The maximum error in phase we introduce by neglecting $v_\phi$ is therefore 
\begin{equation}
  \xi_{v_\phi}-\xi_c = k(v_\phi - c)t \lesssim 0.6. \label{eqn:dxi-calc}
\end{equation}
Solving for the phase velocity gives us $v_\phi/c - 1 < 2\times 10^{-3}$.

On the basis of this calculation and the good agreement between calculated phase and $W_y$ seen for the electrons (Figure \ref{fig:sample-traj}, and others), we therefore consider it suitable to neglect the difference between the phase velocity and the speed of light in this work.

\section{Robustness of two-stage DLA model to alternate parameters} 
\label{append:other-parameters}

\begin{figure*}
    \centering
    \begin{subfigure}{0.47\textwidth}
    \includegraphics[width=0.95\linewidth]{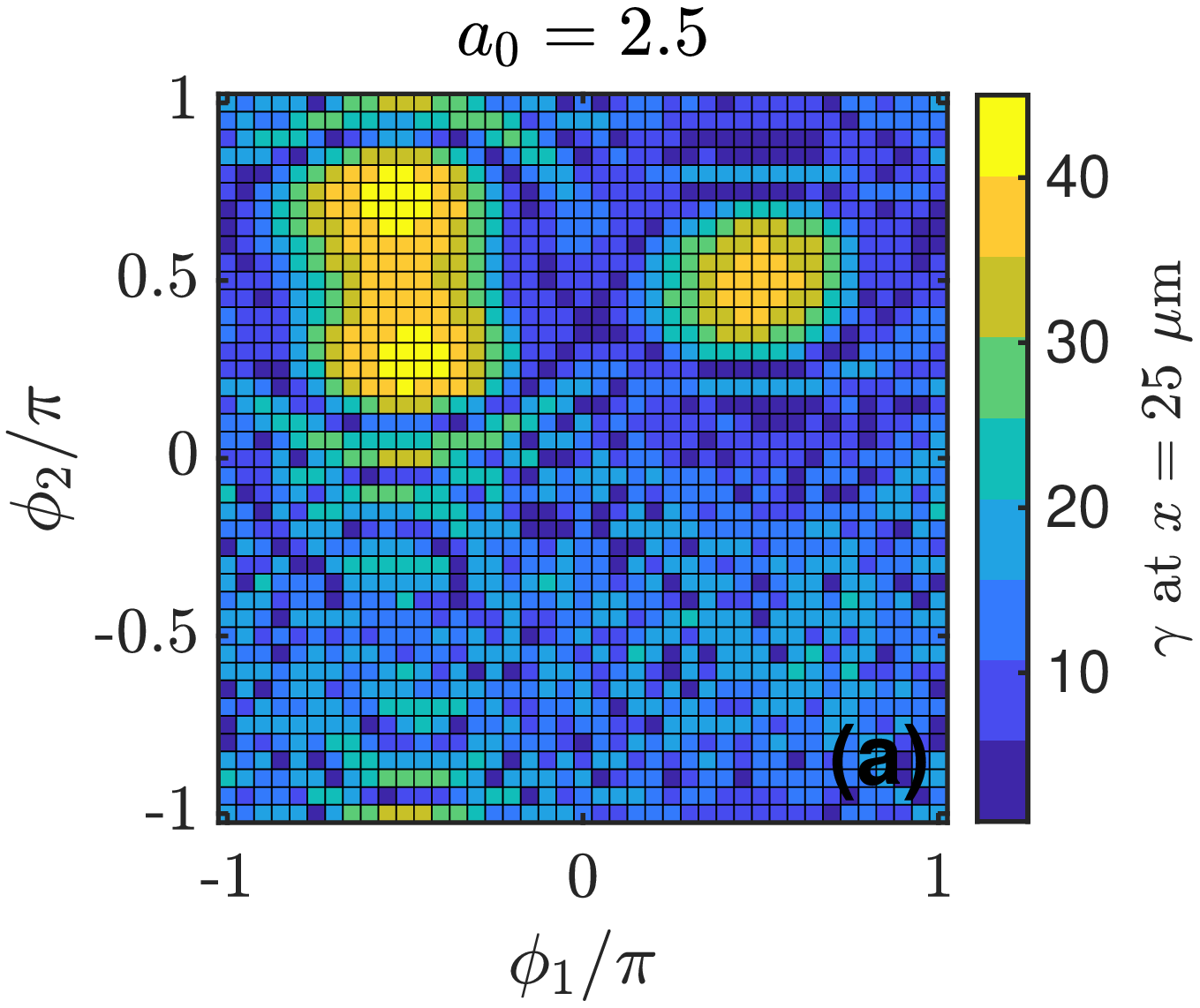}
    \subcaption{} \label{fig:plane-g-phase-lower_a0}
    \end{subfigure}    
    \begin{subfigure}{0.47\textwidth}
    \includegraphics[width=0.95\linewidth]{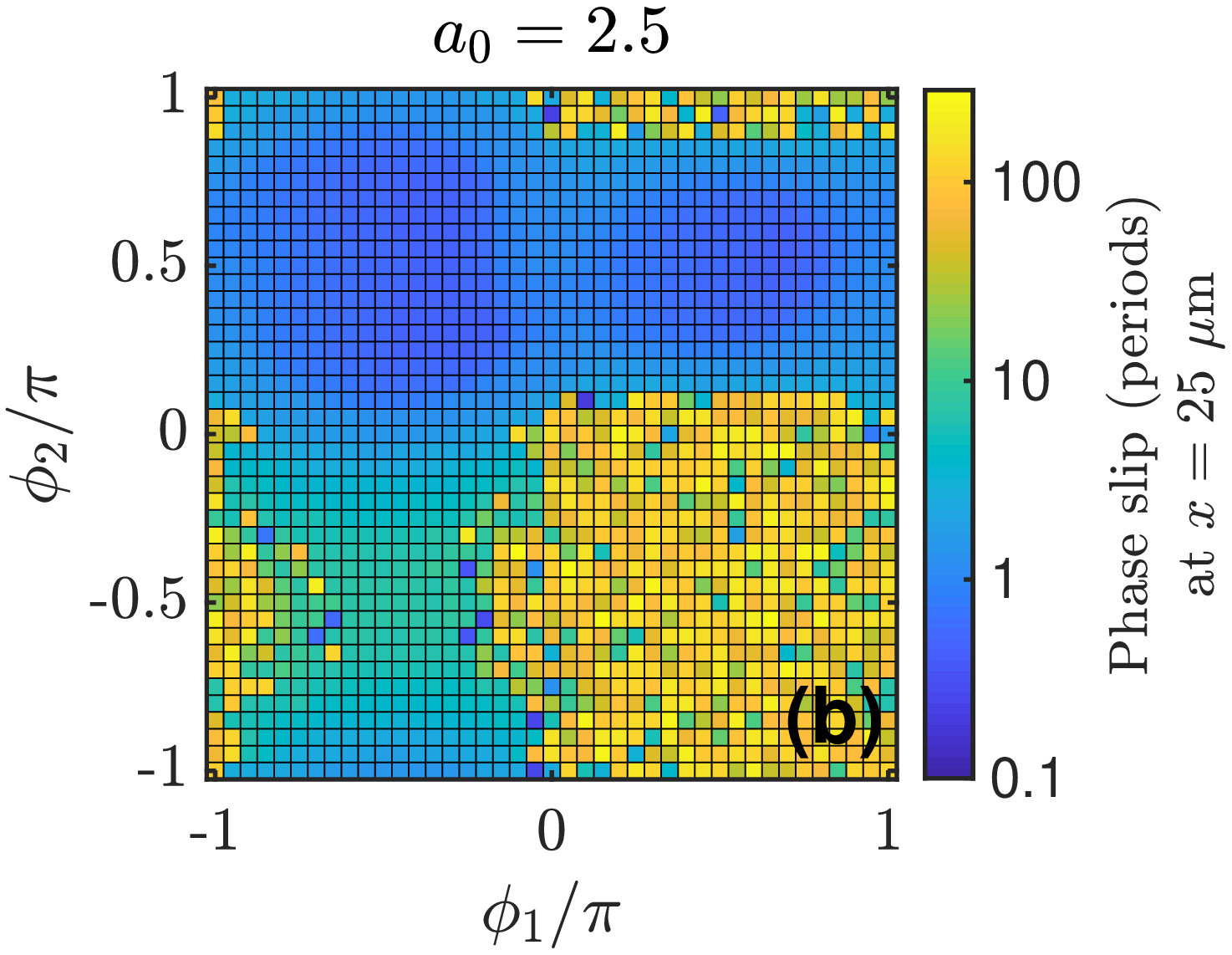}
    \subcaption{} \label{fig:plane-ph-phase-lower_a0}
    \end{subfigure}
    \vspace{-1.75\baselineskip}
    \caption{
    Evaluation of electron trajectories in counter-propagating lasers for $a_0=2.5$, shelf length $x=25$ $\mu$m, and initial longitudinal momentum $p_{x,0}/mc=5$ and transverse momentum $p_{y,0}/mc=2.5$. Akin to Figure \ref{fig:plane-phase}, we scan over the initial laser phases $\phi_1$ and $\phi_2$. (a) $\gamma$-factor after propagation over the shelf length and (b) corresponding phase slip of the electron in the $+x$-propagating laser.
    }
    \label{fig:plane-phase-lower_a0}
\end{figure*}

In this Appendix, we demonstrate that the two-stage acceleration process we identify in simulations remains conceptually valid for an alternate choice of parameters, which suggests the process may be robust to changes in the shelf length and laser intensity. Figure \ref{fig:plane-phase-lower_a0} shows the result of repeating the scan over initial phases done to produce Figure \ref{fig:plane-phase} with $a_0 = 2.5$, where we have also reduced the initial electron momenta and the shelf length by a factor of 2. 

We can see from Figure \ref{fig:plane-phase-lower_a0} that the key requirements can still be satisfied for the initial stochastic backward acceleration stage to enable a subsequent non-stochastic, higher energy forward acceleration stage. In particular, we observe that:
\begin{enumerate}
    \item The regular regime, which we can easily identify by the lack of sensitivity to initial conditions seen in Figure \ref{fig:plane-ph-phase-lower_a0} produces higher peak energy over the shelf length than the stochastic regime for sufficient pre-acceleration (for instance, the pre-acceleration $p_{x,0}/mc = 5$ used to generate Figure \ref{fig:plane-phase-lower_a0}).
    \item The stochastic regime can produce high enough energy to enable the regular acceleration (the peak stochastic energy gain is substantially above the initial $p_{x,0}$).
\end{enumerate}

While there are also additional considerations which could be important in the context of a full PIC simulation, our semi-analytic model calculation shows that the acceleration process we observe under our original conditions is still conceptually valid for these new conditions.


\begin{thebibliography}{49}%
\makeatletter
\providecommand \@ifxundefined [1]{%
 \@ifx{#1\undefined}
}%
\providecommand \@ifnum [1]{%
 \ifnum #1\expandafter \@firstoftwo
 \else \expandafter \@secondoftwo
 \fi
}%
\providecommand \@ifx [1]{%
 \ifx #1\expandafter \@firstoftwo
 \else \expandafter \@secondoftwo
 \fi
}%
\providecommand \natexlab [1]{#1}%
\providecommand \enquote  [1]{``#1''}%
\providecommand \bibnamefont  [1]{#1}%
\providecommand \bibfnamefont [1]{#1}%
\providecommand \citenamefont [1]{#1}%
\providecommand \href@noop [0]{\@secondoftwo}%
\providecommand \href [0]{\begingroup \@sanitize@url \@href}%
\providecommand \@href[1]{\@@startlink{#1}\@@href}%
\providecommand \@@href[1]{\endgroup#1\@@endlink}%
\providecommand \@sanitize@url [0]{\catcode `\\12\catcode `\$12\catcode
  `\&12\catcode `\#12\catcode `\^12\catcode `\_12\catcode `\%12\relax}%
\providecommand \@@startlink[1]{}%
\providecommand \@@endlink[0]{}%
\providecommand \url  [0]{\begingroup\@sanitize@url \@url }%
\providecommand \@url [1]{\endgroup\@href {#1}{\urlprefix }}%
\providecommand \urlprefix  [0]{URL }%
\providecommand \Eprint [0]{\href }%
\providecommand \doibase [0]{http://dx.doi.org/}%
\providecommand \selectlanguage [0]{\@gobble}%
\providecommand \bibinfo  [0]{\@secondoftwo}%
\providecommand \bibfield  [0]{\@secondoftwo}%
\providecommand \translation [1]{[#1]}%
\providecommand \BibitemOpen [0]{}%
\providecommand \bibitemStop [0]{}%
\providecommand \bibitemNoStop [0]{.\EOS\space}%
\providecommand \EOS [0]{\spacefactor3000\relax}%
\providecommand \BibitemShut  [1]{\csname bibitem#1\endcsname}%
\let\auto@bib@innerbib\@empty
\bibitem [{\citenamefont {Tajima}\ and\ \citenamefont
  {Dawson}(1979{\natexlab{a}})}]{tajima1979}%
  \BibitemOpen
  \bibfield  {author} {\bibinfo {author} {\bibfnamefont {T.}~\bibnamefont
  {Tajima}}\ and\ \bibinfo {author} {\bibfnamefont {J.~M.}\ \bibnamefont
  {Dawson}},\ }\href {\doibase 10.1103/PhysRevLett.43.267} {\bibfield
  {journal} {\bibinfo  {journal} {Phys. Rev. Lett.}\ }\textbf {\bibinfo
  {volume} {43}},\ \bibinfo {pages} {267} (\bibinfo {year}
  {1979}{\natexlab{a}})}\BibitemShut {NoStop}%
\bibitem [{\citenamefont {Malka}(2012)}]{malka2012accelerators_review}%
  \BibitemOpen
  \bibfield  {author} {\bibinfo {author} {\bibfnamefont {V.}~\bibnamefont
  {Malka}},\ }\href {\doibase 10.1063/1.3695389} {\bibfield  {journal}
  {\bibinfo  {journal} {Physics of Plasmas}\ }\textbf {\bibinfo {volume}
  {19}},\ \bibinfo {pages} {055501} (\bibinfo {year} {2012})}\BibitemShut
  {NoStop}%
\bibitem [{\citenamefont {Li}\ \emph {et~al.}(2006)\citenamefont {Li},
  \citenamefont {S\'eguin}, \citenamefont {Frenje}, \citenamefont {Rygg},
  \citenamefont {Petrasso}, \citenamefont {Town}, \citenamefont {Amendt},
  \citenamefont {Hatchett}, \citenamefont {Landen}, \citenamefont {Mackinnon},
  \citenamefont {Patel}, \citenamefont {Smalyuk}, \citenamefont {Sangster},\
  and\ \citenamefont {Knauer}}]{li2006proton_radiography}%
  \BibitemOpen
  \bibfield  {author} {\bibinfo {author} {\bibfnamefont {C.~K.}\ \bibnamefont
  {Li}}, \bibinfo {author} {\bibfnamefont {F.~H.}\ \bibnamefont {S\'eguin}},
  \bibinfo {author} {\bibfnamefont {J.~A.}\ \bibnamefont {Frenje}}, \bibinfo
  {author} {\bibfnamefont {J.~R.}\ \bibnamefont {Rygg}}, \bibinfo {author}
  {\bibfnamefont {R.~D.}\ \bibnamefont {Petrasso}}, \bibinfo {author}
  {\bibfnamefont {R.~P.~J.}\ \bibnamefont {Town}}, \bibinfo {author}
  {\bibfnamefont {P.~A.}\ \bibnamefont {Amendt}}, \bibinfo {author}
  {\bibfnamefont {S.~P.}\ \bibnamefont {Hatchett}}, \bibinfo {author}
  {\bibfnamefont {O.~L.}\ \bibnamefont {Landen}}, \bibinfo {author}
  {\bibfnamefont {A.~J.}\ \bibnamefont {Mackinnon}}, \bibinfo {author}
  {\bibfnamefont {P.~K.}\ \bibnamefont {Patel}}, \bibinfo {author}
  {\bibfnamefont {V.~A.}\ \bibnamefont {Smalyuk}}, \bibinfo {author}
  {\bibfnamefont {T.~C.}\ \bibnamefont {Sangster}}, \ and\ \bibinfo {author}
  {\bibfnamefont {J.~P.}\ \bibnamefont {Knauer}},\ }\href {\doibase
  10.1103/PhysRevLett.97.135003} {\bibfield  {journal} {\bibinfo  {journal}
  {Phys. Rev. Lett.}\ }\textbf {\bibinfo {volume} {97}},\ \bibinfo {pages}
  {135003} (\bibinfo {year} {2006})}\BibitemShut {NoStop}%
\bibitem [{\citenamefont {Kneip}\ \emph {et~al.}(2011)\citenamefont {Kneip},
  \citenamefont {McGuffey}, \citenamefont {Dollar}, \citenamefont {Bloom},
  \citenamefont {Chvykov}, \citenamefont {Kalintchenko}, \citenamefont
  {Krushelnick}, \citenamefont {Maksimchuk}, \citenamefont {Mangles},
  \citenamefont {Matsuoka}, \citenamefont {Najmudin}, \citenamefont {Palmer},
  \citenamefont {Schreiber}, \citenamefont {Schumaker}, \citenamefont
  {Thomas},\ and\ \citenamefont {Yanovsky}}]{kneip2011frog}%
  \BibitemOpen
  \bibfield  {author} {\bibinfo {author} {\bibfnamefont {S.}~\bibnamefont
  {Kneip}}, \bibinfo {author} {\bibfnamefont {C.}~\bibnamefont {McGuffey}},
  \bibinfo {author} {\bibfnamefont {F.}~\bibnamefont {Dollar}}, \bibinfo
  {author} {\bibfnamefont {M.~S.}\ \bibnamefont {Bloom}}, \bibinfo {author}
  {\bibfnamefont {V.}~\bibnamefont {Chvykov}}, \bibinfo {author} {\bibfnamefont
  {G.}~\bibnamefont {Kalintchenko}}, \bibinfo {author} {\bibfnamefont
  {K.}~\bibnamefont {Krushelnick}}, \bibinfo {author} {\bibfnamefont
  {A.}~\bibnamefont {Maksimchuk}}, \bibinfo {author} {\bibfnamefont {S.~P.~D.}\
  \bibnamefont {Mangles}}, \bibinfo {author} {\bibfnamefont {T.}~\bibnamefont
  {Matsuoka}}, \bibinfo {author} {\bibfnamefont {Z.}~\bibnamefont {Najmudin}},
  \bibinfo {author} {\bibfnamefont {C.~A.~J.}\ \bibnamefont {Palmer}}, \bibinfo
  {author} {\bibfnamefont {J.}~\bibnamefont {Schreiber}}, \bibinfo {author}
  {\bibfnamefont {W.}~\bibnamefont {Schumaker}}, \bibinfo {author}
  {\bibfnamefont {A.~G.~R.}\ \bibnamefont {Thomas}}, \ and\ \bibinfo {author}
  {\bibfnamefont {V.}~\bibnamefont {Yanovsky}},\ }\href {\doibase
  10.1063/1.3627216} {\bibfield  {journal} {\bibinfo  {journal} {Applied
  Physics Letters}\ }\textbf {\bibinfo {volume} {99}},\ \bibinfo {pages}
  {093701} (\bibinfo {year} {2011})}\BibitemShut {NoStop}%
\bibitem [{\citenamefont {Bulanov}\ \emph {et~al.}(2015)\citenamefont
  {Bulanov}, \citenamefont {Esirkepov}, \citenamefont {Kando}, \citenamefont
  {Koga}, \citenamefont {Kondo},\ and\ \citenamefont
  {Korn}}]{bulanov2015labastro}%
  \BibitemOpen
  \bibfield  {author} {\bibinfo {author} {\bibfnamefont {S.~V.}\ \bibnamefont
  {Bulanov}}, \bibinfo {author} {\bibfnamefont {T.~Z.}\ \bibnamefont
  {Esirkepov}}, \bibinfo {author} {\bibfnamefont {M.}~\bibnamefont {Kando}},
  \bibinfo {author} {\bibfnamefont {J.}~\bibnamefont {Koga}}, \bibinfo {author}
  {\bibfnamefont {K.}~\bibnamefont {Kondo}}, \ and\ \bibinfo {author}
  {\bibfnamefont {G.}~\bibnamefont {Korn}},\ }\href {\doibase
  10.1134/S1063780X15010018} {\bibfield  {journal} {\bibinfo  {journal} {Plasma
  Physics Reports}\ }\textbf {\bibinfo {volume} {41}},\ \bibinfo {pages} {1}
  (\bibinfo {year} {2015})}\BibitemShut {NoStop}%
\bibitem [{\citenamefont {Lindl}(1995)}]{lindl1995icf}%
  \BibitemOpen
  \bibfield  {author} {\bibinfo {author} {\bibfnamefont {J.}~\bibnamefont
  {Lindl}},\ }\href {\doibase 10.1063/1.871025} {\bibfield  {journal} {\bibinfo
   {journal} {Physics of Plasmas}\ }\textbf {\bibinfo {volume} {2}},\ \bibinfo
  {pages} {3933} (\bibinfo {year} {1995})}\BibitemShut {NoStop}%
\bibitem [{\citenamefont {Roth}\ \emph {et~al.}(2001)\citenamefont {Roth},
  \citenamefont {Cowan}, \citenamefont {Key}, \citenamefont {Hatchett},
  \citenamefont {Brown}, \citenamefont {Fountain}, \citenamefont {Johnson},
  \citenamefont {Pennington}, \citenamefont {Snavely}, \citenamefont {Wilks},
  \citenamefont {Yasuike}, \citenamefont {Ruhl}, \citenamefont {Pegoraro},
  \citenamefont {Bulanov}, \citenamefont {Campbell}, \citenamefont {Perry},\
  and\ \citenamefont {Powell}}]{roth2001proton_icf}%
  \BibitemOpen
  \bibfield  {author} {\bibinfo {author} {\bibfnamefont {M.}~\bibnamefont
  {Roth}}, \bibinfo {author} {\bibfnamefont {T.~E.}\ \bibnamefont {Cowan}},
  \bibinfo {author} {\bibfnamefont {M.~H.}\ \bibnamefont {Key}}, \bibinfo
  {author} {\bibfnamefont {S.~P.}\ \bibnamefont {Hatchett}}, \bibinfo {author}
  {\bibfnamefont {C.}~\bibnamefont {Brown}}, \bibinfo {author} {\bibfnamefont
  {W.}~\bibnamefont {Fountain}}, \bibinfo {author} {\bibfnamefont
  {J.}~\bibnamefont {Johnson}}, \bibinfo {author} {\bibfnamefont {D.~M.}\
  \bibnamefont {Pennington}}, \bibinfo {author} {\bibfnamefont {R.~A.}\
  \bibnamefont {Snavely}}, \bibinfo {author} {\bibfnamefont {S.~C.}\
  \bibnamefont {Wilks}}, \bibinfo {author} {\bibfnamefont {K.}~\bibnamefont
  {Yasuike}}, \bibinfo {author} {\bibfnamefont {H.}~\bibnamefont {Ruhl}},
  \bibinfo {author} {\bibfnamefont {F.}~\bibnamefont {Pegoraro}}, \bibinfo
  {author} {\bibfnamefont {S.~V.}\ \bibnamefont {Bulanov}}, \bibinfo {author}
  {\bibfnamefont {E.~M.}\ \bibnamefont {Campbell}}, \bibinfo {author}
  {\bibfnamefont {M.~D.}\ \bibnamefont {Perry}}, \ and\ \bibinfo {author}
  {\bibfnamefont {H.}~\bibnamefont {Powell}},\ }\href {\doibase
  10.1103/PhysRevLett.86.436} {\bibfield  {journal} {\bibinfo  {journal} {Phys.
  Rev. Lett.}\ }\textbf {\bibinfo {volume} {86}},\ \bibinfo {pages} {436}
  (\bibinfo {year} {2001})}\BibitemShut {NoStop}%
\bibitem [{\citenamefont {Macchi}, \citenamefont {Borghesi},\ and\
  \citenamefont {Passoni}(2013)}]{macchi2013ion_accel_rev}%
  \BibitemOpen
  \bibfield  {author} {\bibinfo {author} {\bibfnamefont {A.}~\bibnamefont
  {Macchi}}, \bibinfo {author} {\bibfnamefont {M.}~\bibnamefont {Borghesi}}, \
  and\ \bibinfo {author} {\bibfnamefont {M.}~\bibnamefont {Passoni}},\ }\href
  {\doibase 10.1103/RevModPhys.85.751} {\bibfield  {journal} {\bibinfo
  {journal} {Rev. Mod. Phys.}\ }\textbf {\bibinfo {volume} {85}},\ \bibinfo
  {pages} {751} (\bibinfo {year} {2013})}\BibitemShut {NoStop}%
\bibitem [{\citenamefont {Pomerantz}\ \emph {et~al.}(2014)\citenamefont
  {Pomerantz}, \citenamefont {McCary}, \citenamefont {Meadows}, \citenamefont
  {Arefiev}, \citenamefont {Bernstein}, \citenamefont {Chester}, \citenamefont
  {Cortez}, \citenamefont {Donovan}, \citenamefont {Dyer}, \citenamefont
  {Gaul}, \citenamefont {Hamilton}, \citenamefont {Kuk}, \citenamefont
  {Lestrade}, \citenamefont {Wang}, \citenamefont {Ditmire},\ and\
  \citenamefont {Hegelich}}]{pomerantz2014neutrons}%
  \BibitemOpen
  \bibfield  {author} {\bibinfo {author} {\bibfnamefont {I.}~\bibnamefont
  {Pomerantz}}, \bibinfo {author} {\bibfnamefont {E.}~\bibnamefont {McCary}},
  \bibinfo {author} {\bibfnamefont {A.~R.}\ \bibnamefont {Meadows}}, \bibinfo
  {author} {\bibfnamefont {A.}~\bibnamefont {Arefiev}}, \bibinfo {author}
  {\bibfnamefont {A.~C.}\ \bibnamefont {Bernstein}}, \bibinfo {author}
  {\bibfnamefont {C.}~\bibnamefont {Chester}}, \bibinfo {author} {\bibfnamefont
  {J.}~\bibnamefont {Cortez}}, \bibinfo {author} {\bibfnamefont {M.~E.}\
  \bibnamefont {Donovan}}, \bibinfo {author} {\bibfnamefont {G.}~\bibnamefont
  {Dyer}}, \bibinfo {author} {\bibfnamefont {E.~W.}\ \bibnamefont {Gaul}},
  \bibinfo {author} {\bibfnamefont {D.}~\bibnamefont {Hamilton}}, \bibinfo
  {author} {\bibfnamefont {D.}~\bibnamefont {Kuk}}, \bibinfo {author}
  {\bibfnamefont {A.~C.}\ \bibnamefont {Lestrade}}, \bibinfo {author}
  {\bibfnamefont {C.}~\bibnamefont {Wang}}, \bibinfo {author} {\bibfnamefont
  {T.}~\bibnamefont {Ditmire}}, \ and\ \bibinfo {author} {\bibfnamefont
  {B.~M.}\ \bibnamefont {Hegelich}},\ }\href {\doibase
  10.1103/PhysRevLett.113.184801} {\bibfield  {journal} {\bibinfo  {journal}
  {Phys. Rev. Lett.}\ }\textbf {\bibinfo {volume} {113}},\ \bibinfo {pages}
  {184801} (\bibinfo {year} {2014})}\BibitemShut {NoStop}%
\bibitem [{\citenamefont {Kmetec}\ \emph {et~al.}(1992)\citenamefont {Kmetec},
  \citenamefont {Gordon}, \citenamefont {Macklin}, \citenamefont {Lemoff},
  \citenamefont {Brown},\ and\ \citenamefont {Harris}}]{kmetec1992brem}%
  \BibitemOpen
  \bibfield  {author} {\bibinfo {author} {\bibfnamefont {J.~D.}\ \bibnamefont
  {Kmetec}}, \bibinfo {author} {\bibfnamefont {C.~L.}\ \bibnamefont {Gordon}},
  \bibinfo {author} {\bibfnamefont {J.~J.}\ \bibnamefont {Macklin}}, \bibinfo
  {author} {\bibfnamefont {B.~E.}\ \bibnamefont {Lemoff}}, \bibinfo {author}
  {\bibfnamefont {G.~S.}\ \bibnamefont {Brown}}, \ and\ \bibinfo {author}
  {\bibfnamefont {S.~E.}\ \bibnamefont {Harris}},\ }\href {\doibase
  10.1103/PhysRevLett.68.1527} {\bibfield  {journal} {\bibinfo  {journal}
  {Phys. Rev. Lett.}\ }\textbf {\bibinfo {volume} {68}},\ \bibinfo {pages}
  {1527} (\bibinfo {year} {1992})}\BibitemShut {NoStop}%
\bibitem [{\citenamefont {Stark}, \citenamefont {Toncian},\ and\ \citenamefont
  {Arefiev}(2016)}]{stark2016xray}%
  \BibitemOpen
  \bibfield  {author} {\bibinfo {author} {\bibfnamefont {D.~J.}\ \bibnamefont
  {Stark}}, \bibinfo {author} {\bibfnamefont {T.}~\bibnamefont {Toncian}}, \
  and\ \bibinfo {author} {\bibfnamefont {A.~V.}\ \bibnamefont {Arefiev}},\
  }\href {\doibase 10.1103/PhysRevLett.116.185003} {\bibfield  {journal}
  {\bibinfo  {journal} {Phys. Rev. Lett.}\ }\textbf {\bibinfo {volume} {116}},\
  \bibinfo {pages} {185003} (\bibinfo {year} {2016})}\BibitemShut {NoStop}%
\bibitem [{\citenamefont {Hartemann}\ \emph {et~al.}(1995)\citenamefont
  {Hartemann}, \citenamefont {Fochs}, \citenamefont {Le~Sage}, \citenamefont
  {Luhmann}, \citenamefont {Woodworth}, \citenamefont {Perry}, \citenamefont
  {Chen},\ and\ \citenamefont {Kerman}}]{hartemann1995dla}%
  \BibitemOpen
  \bibfield  {author} {\bibinfo {author} {\bibfnamefont {F.~V.}\ \bibnamefont
  {Hartemann}}, \bibinfo {author} {\bibfnamefont {S.~N.}\ \bibnamefont
  {Fochs}}, \bibinfo {author} {\bibfnamefont {G.~P.}\ \bibnamefont {Le~Sage}},
  \bibinfo {author} {\bibfnamefont {N.~C.}\ \bibnamefont {Luhmann}}, \bibinfo
  {author} {\bibfnamefont {J.~G.}\ \bibnamefont {Woodworth}}, \bibinfo {author}
  {\bibfnamefont {M.~D.}\ \bibnamefont {Perry}}, \bibinfo {author}
  {\bibfnamefont {Y.~J.}\ \bibnamefont {Chen}}, \ and\ \bibinfo {author}
  {\bibfnamefont {A.~K.}\ \bibnamefont {Kerman}},\ }\href {\doibase
  10.1103/PhysRevE.51.4833} {\bibfield  {journal} {\bibinfo  {journal} {Phys.
  Rev. E}\ }\textbf {\bibinfo {volume} {51}},\ \bibinfo {pages} {4833}
  (\bibinfo {year} {1995})}\BibitemShut {NoStop}%
\bibitem [{\citenamefont {Tajima}\ and\ \citenamefont
  {Dawson}(1979{\natexlab{b}})}]{tajima1979laser}%
  \BibitemOpen
  \bibfield  {author} {\bibinfo {author} {\bibfnamefont {T.}~\bibnamefont
  {Tajima}}\ and\ \bibinfo {author} {\bibfnamefont {J.~M.}\ \bibnamefont
  {Dawson}},\ }\href@noop {} {\bibfield  {journal} {\bibinfo  {journal}
  {Physical Review Letters}\ }\textbf {\bibinfo {volume} {43}},\ \bibinfo
  {pages} {267} (\bibinfo {year} {1979}{\natexlab{b}})}\BibitemShut {NoStop}%
\bibitem [{\citenamefont {Kaw}\ and\ \citenamefont
  {Dawson}(1970)}]{kaw1970rel_trans}%
  \BibitemOpen
  \bibfield  {author} {\bibinfo {author} {\bibfnamefont {P.}~\bibnamefont
  {Kaw}}\ and\ \bibinfo {author} {\bibfnamefont {J.}~\bibnamefont {Dawson}},\
  }\href {\doibase 10.1063/1.1692942} {\bibfield  {journal} {\bibinfo
  {journal} {The Physics of Fluids}\ }\textbf {\bibinfo {volume} {13}},\
  \bibinfo {pages} {472} (\bibinfo {year} {1970})}\BibitemShut {NoStop}%
\bibitem [{\citenamefont {Kruer}\ and\ \citenamefont
  {Estabrook}(1985)}]{kruer1985jxB}%
  \BibitemOpen
  \bibfield  {author} {\bibinfo {author} {\bibfnamefont {W.~L.}\ \bibnamefont
  {Kruer}}\ and\ \bibinfo {author} {\bibfnamefont {K.}~\bibnamefont
  {Estabrook}},\ }\href {\doibase 10.1063/1.865171} {\bibfield  {journal}
  {\bibinfo  {journal} {The Physics of Fluids}\ }\textbf {\bibinfo {volume}
  {28}},\ \bibinfo {pages} {430} (\bibinfo {year} {1985})}\BibitemShut
  {NoStop}%
\bibitem [{\citenamefont {Lefebvre}\ and\ \citenamefont
  {Bonnaud}(1997)}]{lefebvre1997absorption}%
  \BibitemOpen
  \bibfield  {author} {\bibinfo {author} {\bibfnamefont {E.}~\bibnamefont
  {Lefebvre}}\ and\ \bibinfo {author} {\bibfnamefont {G.}~\bibnamefont
  {Bonnaud}},\ }\href {\doibase 10.1103/PhysRevE.55.1011} {\bibfield  {journal}
  {\bibinfo  {journal} {Phys. Rev. E}\ }\textbf {\bibinfo {volume} {55}},\
  \bibinfo {pages} {1011} (\bibinfo {year} {1997})}\BibitemShut {NoStop}%
\bibitem [{\citenamefont {May}\ \emph {et~al.}(2011)\citenamefont {May},
  \citenamefont {Tonge}, \citenamefont {Fiuza}, \citenamefont {Fonseca},
  \citenamefont {Silva}, \citenamefont {Ren},\ and\ \citenamefont
  {Mori}}]{may2011absorption}%
  \BibitemOpen
  \bibfield  {author} {\bibinfo {author} {\bibfnamefont {J.}~\bibnamefont
  {May}}, \bibinfo {author} {\bibfnamefont {J.}~\bibnamefont {Tonge}}, \bibinfo
  {author} {\bibfnamefont {F.}~\bibnamefont {Fiuza}}, \bibinfo {author}
  {\bibfnamefont {R.~A.}\ \bibnamefont {Fonseca}}, \bibinfo {author}
  {\bibfnamefont {L.~O.}\ \bibnamefont {Silva}}, \bibinfo {author}
  {\bibfnamefont {C.}~\bibnamefont {Ren}}, \ and\ \bibinfo {author}
  {\bibfnamefont {W.~B.}\ \bibnamefont {Mori}},\ }\href {\doibase
  10.1103/PhysRevE.84.025401} {\bibfield  {journal} {\bibinfo  {journal} {Phys.
  Rev. E}\ }\textbf {\bibinfo {volume} {84}},\ \bibinfo {pages} {025401}
  (\bibinfo {year} {2011})}\BibitemShut {NoStop}%
\bibitem [{\citenamefont {Wilks}\ \emph {et~al.}(1992)\citenamefont {Wilks},
  \citenamefont {Kruer}, \citenamefont {Tabak},\ and\ \citenamefont
  {Langdon}}]{wilks1992pond_scale}%
  \BibitemOpen
  \bibfield  {author} {\bibinfo {author} {\bibfnamefont {S.~C.}\ \bibnamefont
  {Wilks}}, \bibinfo {author} {\bibfnamefont {W.~L.}\ \bibnamefont {Kruer}},
  \bibinfo {author} {\bibfnamefont {M.}~\bibnamefont {Tabak}}, \ and\ \bibinfo
  {author} {\bibfnamefont {A.~B.}\ \bibnamefont {Langdon}},\ }\href {\doibase
  10.1103/PhysRevLett.69.1383} {\bibfield  {journal} {\bibinfo  {journal}
  {Phys. Rev. Lett.}\ }\textbf {\bibinfo {volume} {69}},\ \bibinfo {pages}
  {1383} (\bibinfo {year} {1992})}\BibitemShut {NoStop}%
\bibitem [{\citenamefont {Pukhov}, \citenamefont {Sheng},\ and\ \citenamefont
  {Meyer-ter Vehn}(1999)}]{pukhov1999betatron}%
  \BibitemOpen
  \bibfield  {author} {\bibinfo {author} {\bibfnamefont {A.}~\bibnamefont
  {Pukhov}}, \bibinfo {author} {\bibfnamefont {Z.-M.}\ \bibnamefont {Sheng}}, \
  and\ \bibinfo {author} {\bibfnamefont {J.}~\bibnamefont {Meyer-ter Vehn}},\
  }\href {\doibase 10.1063/1.873242} {\bibfield  {journal} {\bibinfo  {journal}
  {Physics of Plasmas}\ }\textbf {\bibinfo {volume} {6}},\ \bibinfo {pages}
  {2847} (\bibinfo {year} {1999})}\BibitemShut {NoStop}%
\bibitem [{\citenamefont {Arefiev}\ \emph {et~al.}(2012)\citenamefont
  {Arefiev}, \citenamefont {Breizman}, \citenamefont {Schollmeier},\ and\
  \citenamefont {Khudik}}]{arefiev2012channel_accel}%
  \BibitemOpen
  \bibfield  {author} {\bibinfo {author} {\bibfnamefont {A.~V.}\ \bibnamefont
  {Arefiev}}, \bibinfo {author} {\bibfnamefont {B.~N.}\ \bibnamefont
  {Breizman}}, \bibinfo {author} {\bibfnamefont {M.}~\bibnamefont
  {Schollmeier}}, \ and\ \bibinfo {author} {\bibfnamefont {V.~N.}\ \bibnamefont
  {Khudik}},\ }\href {\doibase 10.1103/PhysRevLett.108.145004} {\bibfield
  {journal} {\bibinfo  {journal} {Phys. Rev. Lett.}\ }\textbf {\bibinfo
  {volume} {108}},\ \bibinfo {pages} {145004} (\bibinfo {year}
  {2012})}\BibitemShut {NoStop}%
\bibitem [{\citenamefont {Kemp}, \citenamefont {Sentoku},\ and\ \citenamefont
  {Tabak}(2009)}]{kemp2009psshelf_pond}%
  \BibitemOpen
  \bibfield  {author} {\bibinfo {author} {\bibfnamefont {A.~J.}\ \bibnamefont
  {Kemp}}, \bibinfo {author} {\bibfnamefont {Y.}~\bibnamefont {Sentoku}}, \
  and\ \bibinfo {author} {\bibfnamefont {M.}~\bibnamefont {Tabak}},\ }\href
  {\doibase 10.1103/PhysRevE.79.066406} {\bibfield  {journal} {\bibinfo
  {journal} {Phys. Rev. E}\ }\textbf {\bibinfo {volume} {79}},\ \bibinfo
  {pages} {066406} (\bibinfo {year} {2009})}\BibitemShut {NoStop}%
\bibitem [{\citenamefont {Mendon\ifmmode~\mbox{\c{c}}\else
  \c{c}\fi{}a}(1983)}]{mendonca1983stochastic}%
  \BibitemOpen
  \bibfield  {author} {\bibinfo {author} {\bibfnamefont {J.~T.}\ \bibnamefont
  {Mendon\ifmmode~\mbox{\c{c}}\else \c{c}\fi{}a}},\ }\href {\doibase
  10.1103/PhysRevA.28.3592} {\bibfield  {journal} {\bibinfo  {journal} {Phys.
  Rev. A}\ }\textbf {\bibinfo {volume} {28}},\ \bibinfo {pages} {3592}
  (\bibinfo {year} {1983})}\BibitemShut {NoStop}%
\bibitem [{\citenamefont {Rax}(1992)}]{rax1992stochastic_heating}%
  \BibitemOpen
  \bibfield  {author} {\bibinfo {author} {\bibfnamefont {J.~M.}\ \bibnamefont
  {Rax}},\ }\href {\doibase 10.1063/1.860299} {\bibfield  {journal} {\bibinfo
  {journal} {Physics of Fluids B: Plasma Physics}\ }\textbf {\bibinfo {volume}
  {4}},\ \bibinfo {pages} {3962} (\bibinfo {year} {1992})}\BibitemShut
  {NoStop}%
\bibitem [{\citenamefont {Sheng}\ \emph {et~al.}(2002)\citenamefont {Sheng},
  \citenamefont {Mima}, \citenamefont {Sentoku}, \citenamefont
  {Jovanovi\ifmmode~\acute{c}\else \'{c}\fi{}}, \citenamefont {Taguchi},
  \citenamefont {Zhang},\ and\ \citenamefont {Meyer-ter
  Vehn}}]{sheng2002stochastic_theory}%
  \BibitemOpen
  \bibfield  {author} {\bibinfo {author} {\bibfnamefont {Z.-M.}\ \bibnamefont
  {Sheng}}, \bibinfo {author} {\bibfnamefont {K.}~\bibnamefont {Mima}},
  \bibinfo {author} {\bibfnamefont {Y.}~\bibnamefont {Sentoku}}, \bibinfo
  {author} {\bibfnamefont {M.~S.}\ \bibnamefont
  {Jovanovi\ifmmode~\acute{c}\else \'{c}\fi{}}}, \bibinfo {author}
  {\bibfnamefont {T.}~\bibnamefont {Taguchi}}, \bibinfo {author} {\bibfnamefont
  {J.}~\bibnamefont {Zhang}}, \ and\ \bibinfo {author} {\bibfnamefont
  {J.}~\bibnamefont {Meyer-ter Vehn}},\ }\href {\doibase
  10.1103/PhysRevLett.88.055004} {\bibfield  {journal} {\bibinfo  {journal}
  {Phys. Rev. Lett.}\ }\textbf {\bibinfo {volume} {88}},\ \bibinfo {pages}
  {055004} (\bibinfo {year} {2002})}\BibitemShut {NoStop}%
\bibitem [{\citenamefont {Paradkar}\ \emph {et~al.}(2011)\citenamefont
  {Paradkar}, \citenamefont {Wei}, \citenamefont {Yabuuchi}, \citenamefont
  {Stephens}, \citenamefont {Haines}, \citenamefont {Krasheninnikov},\ and\
  \citenamefont {Beg}}]{paradkar2011scalelength}%
  \BibitemOpen
  \bibfield  {author} {\bibinfo {author} {\bibfnamefont {B.~S.}\ \bibnamefont
  {Paradkar}}, \bibinfo {author} {\bibfnamefont {M.~S.}\ \bibnamefont {Wei}},
  \bibinfo {author} {\bibfnamefont {T.}~\bibnamefont {Yabuuchi}}, \bibinfo
  {author} {\bibfnamefont {R.~B.}\ \bibnamefont {Stephens}}, \bibinfo {author}
  {\bibfnamefont {M.~G.}\ \bibnamefont {Haines}}, \bibinfo {author}
  {\bibfnamefont {S.~I.}\ \bibnamefont {Krasheninnikov}}, \ and\ \bibinfo
  {author} {\bibfnamefont {F.~N.}\ \bibnamefont {Beg}},\ }\href {\doibase
  10.1103/PhysRevE.83.046401} {\bibfield  {journal} {\bibinfo  {journal} {Phys.
  Rev. E}\ }\textbf {\bibinfo {volume} {83}},\ \bibinfo {pages} {046401}
  (\bibinfo {year} {2011})}\BibitemShut {NoStop}%
\bibitem [{\citenamefont {Kojima}\ \emph {et~al.}(2019)\citenamefont {Kojima},
  \citenamefont {Hata}, \citenamefont {Iwata}, \citenamefont {Arikawa},
  \citenamefont {Morace}, \citenamefont {Sakata}, \citenamefont {Lee},
  \citenamefont {Matsuo}, \citenamefont {Law}, \citenamefont {Morita} \emph
  {et~al.}}]{kojima2019electromagnetic}%
  \BibitemOpen
  \bibfield  {author} {\bibinfo {author} {\bibfnamefont {S.}~\bibnamefont
  {Kojima}}, \bibinfo {author} {\bibfnamefont {M.}~\bibnamefont {Hata}},
  \bibinfo {author} {\bibfnamefont {N.}~\bibnamefont {Iwata}}, \bibinfo
  {author} {\bibfnamefont {Y.}~\bibnamefont {Arikawa}}, \bibinfo {author}
  {\bibfnamefont {A.}~\bibnamefont {Morace}}, \bibinfo {author} {\bibfnamefont
  {S.}~\bibnamefont {Sakata}}, \bibinfo {author} {\bibfnamefont
  {S.}~\bibnamefont {Lee}}, \bibinfo {author} {\bibfnamefont {K.}~\bibnamefont
  {Matsuo}}, \bibinfo {author} {\bibfnamefont {K.~F.~F.}\ \bibnamefont {Law}},
  \bibinfo {author} {\bibfnamefont {H.}~\bibnamefont {Morita}},  \emph
  {et~al.},\ }\href@noop {} {\bibfield  {journal} {\bibinfo  {journal}
  {Communications Physics}\ }\textbf {\bibinfo {volume} {2}},\ \bibinfo {pages}
  {1} (\bibinfo {year} {2019})}\BibitemShut {NoStop}%
\bibitem [{\citenamefont {Sorokovikova}\ \emph {et~al.}(2016)\citenamefont
  {Sorokovikova}, \citenamefont {Arefiev}, \citenamefont {McGuffey},
  \citenamefont {Qiao}, \citenamefont {Robinson}, \citenamefont {Wei},
  \citenamefont {McLean},\ and\ \citenamefont
  {Beg}}]{sorokovikova2016superpond}%
  \BibitemOpen
  \bibfield  {author} {\bibinfo {author} {\bibfnamefont {A.}~\bibnamefont
  {Sorokovikova}}, \bibinfo {author} {\bibfnamefont {A.~V.}\ \bibnamefont
  {Arefiev}}, \bibinfo {author} {\bibfnamefont {C.}~\bibnamefont {McGuffey}},
  \bibinfo {author} {\bibfnamefont {B.}~\bibnamefont {Qiao}}, \bibinfo {author}
  {\bibfnamefont {A.~P.~L.}\ \bibnamefont {Robinson}}, \bibinfo {author}
  {\bibfnamefont {M.~S.}\ \bibnamefont {Wei}}, \bibinfo {author} {\bibfnamefont
  {H.~S.}\ \bibnamefont {McLean}}, \ and\ \bibinfo {author} {\bibfnamefont
  {F.~N.}\ \bibnamefont {Beg}},\ }\href {\doibase
  10.1103/PhysRevLett.116.155001} {\bibfield  {journal} {\bibinfo  {journal}
  {Phys. Rev. Lett.}\ }\textbf {\bibinfo {volume} {116}},\ \bibinfo {pages}
  {155001} (\bibinfo {year} {2016})}\BibitemShut {NoStop}%
\bibitem [{\citenamefont {Iwata}\ \emph {et~al.}(2018)\citenamefont {Iwata},
  \citenamefont {Kojima}, \citenamefont {Sentoku}, \citenamefont {Hata},\ and\
  \citenamefont {Mima}}]{iwata2018hole_boring}%
  \BibitemOpen
  \bibfield  {author} {\bibinfo {author} {\bibfnamefont {N.}~\bibnamefont
  {Iwata}}, \bibinfo {author} {\bibfnamefont {S.}~\bibnamefont {Kojima}},
  \bibinfo {author} {\bibfnamefont {Y.}~\bibnamefont {Sentoku}}, \bibinfo
  {author} {\bibfnamefont {M.}~\bibnamefont {Hata}}, \ and\ \bibinfo {author}
  {\bibfnamefont {K.}~\bibnamefont {Mima}},\ }\href@noop {} {\bibfield
  {journal} {\bibinfo  {journal} {Nature communications}\ }\textbf {\bibinfo
  {volume} {9}},\ \bibinfo {pages} {623} (\bibinfo {year} {2018})}\BibitemShut
  {NoStop}%
\bibitem [{\citenamefont {{N. Miyanaga}}\ \emph {et~al.}(2006)\citenamefont
  {{N. Miyanaga}}, \citenamefont {{H. Azechi}}, \citenamefont {{K.A. Tanaka}},
  \citenamefont {{T. Kanabe}}, \citenamefont {{T. Jitsuno}}, \citenamefont {{J.
  Kawanaka}}, \citenamefont {{Y. Fujimoto}}, \citenamefont {{R. Kodama}},
  \citenamefont {{H. Shiraga}}, \citenamefont {{K. Knodo}}, \citenamefont {{K.
  Tsubakimoto}}, \citenamefont {{H. Habara}}, \citenamefont {{J.Lu}},
  \citenamefont {{G. Xu}}, \citenamefont {{N. Morio}}, \citenamefont {{S.
  Matsuo}}, \citenamefont {{E. Miyaji}}, \citenamefont {{Y. Kawakami}},
  \citenamefont {{Y. Izawa}},\ and\ \citenamefont {{K.
  Mima}}}]{Miyanaga2006LFEX}%
  \BibitemOpen
  \bibfield  {author} {\bibinfo {author} {\bibnamefont {{N. Miyanaga}}},
  \bibinfo {author} {\bibnamefont {{H. Azechi}}}, \bibinfo {author}
  {\bibnamefont {{K.A. Tanaka}}}, \bibinfo {author} {\bibnamefont {{T.
  Kanabe}}}, \bibinfo {author} {\bibnamefont {{T. Jitsuno}}}, \bibinfo {author}
  {\bibnamefont {{J. Kawanaka}}}, \bibinfo {author} {\bibnamefont {{Y.
  Fujimoto}}}, \bibinfo {author} {\bibnamefont {{R. Kodama}}}, \bibinfo
  {author} {\bibnamefont {{H. Shiraga}}}, \bibinfo {author} {\bibnamefont {{K.
  Knodo}}}, \bibinfo {author} {\bibnamefont {{K. Tsubakimoto}}}, \bibinfo
  {author} {\bibnamefont {{H. Habara}}}, \bibinfo {author} {\bibnamefont
  {{J.Lu}}}, \bibinfo {author} {\bibnamefont {{G. Xu}}}, \bibinfo {author}
  {\bibnamefont {{N. Morio}}}, \bibinfo {author} {\bibnamefont {{S. Matsuo}}},
  \bibinfo {author} {\bibnamefont {{E. Miyaji}}}, \bibinfo {author}
  {\bibnamefont {{Y. Kawakami}}}, \bibinfo {author} {\bibnamefont {{Y.
  Izawa}}}, \ and\ \bibinfo {author} {\bibnamefont {{K. Mima}}},\ }\href
  {\doibase 10.1051/jp4:2006133016} {\bibfield  {journal} {\bibinfo  {journal}
  {J. Phys. IV France}\ }\textbf {\bibinfo {volume} {133}},\ \bibinfo {pages}
  {81} (\bibinfo {year} {2006})}\BibitemShut {NoStop}%
\bibitem [{\citenamefont {Batani}\ \emph {et~al.}(2014)\citenamefont {Batani},
  \citenamefont {Koenig}, \citenamefont {Miquel}, \citenamefont {Ducret},
  \citenamefont {d'Humieres}, \citenamefont {Hulin}, \citenamefont {Caron},
  \citenamefont {Feugeas}, \citenamefont {Nicolai}, \citenamefont {Tikhonchuk},
  \citenamefont {Serani}, \citenamefont {Blanchot}, \citenamefont {Raffestin},
  \citenamefont {Thfoin-Lantuejoul}, \citenamefont {Rosse}, \citenamefont
  {Reverdin}, \citenamefont {Duval}, \citenamefont {Laniesse}, \citenamefont
  {Chanc{\'{e}}}, \citenamefont {Dubreuil}, \citenamefont {Gastineau},
  \citenamefont {Guillard}, \citenamefont {Harrault}, \citenamefont
  {Leb{\oe}uf}, \citenamefont {Ster}, \citenamefont {P{\`{e}}s}, \citenamefont
  {Toussaint}, \citenamefont {Leboeuf}, \citenamefont {Lecherbourg},
  \citenamefont {Szabo}, \citenamefont {Dubois},\ and\ \citenamefont
  {Lubrano-Lavaderci}}]{batani2014PETAL}%
  \BibitemOpen
  \bibfield  {author} {\bibinfo {author} {\bibfnamefont {D.}~\bibnamefont
  {Batani}}, \bibinfo {author} {\bibfnamefont {M.}~\bibnamefont {Koenig}},
  \bibinfo {author} {\bibfnamefont {J.~L.}\ \bibnamefont {Miquel}}, \bibinfo
  {author} {\bibfnamefont {J.~E.}\ \bibnamefont {Ducret}}, \bibinfo {author}
  {\bibfnamefont {E.}~\bibnamefont {d'Humieres}}, \bibinfo {author}
  {\bibfnamefont {S.}~\bibnamefont {Hulin}}, \bibinfo {author} {\bibfnamefont
  {J.}~\bibnamefont {Caron}}, \bibinfo {author} {\bibfnamefont {J.~L.}\
  \bibnamefont {Feugeas}}, \bibinfo {author} {\bibfnamefont {P.}~\bibnamefont
  {Nicolai}}, \bibinfo {author} {\bibfnamefont {V.}~\bibnamefont {Tikhonchuk}},
  \bibinfo {author} {\bibfnamefont {L.}~\bibnamefont {Serani}}, \bibinfo
  {author} {\bibfnamefont {N.}~\bibnamefont {Blanchot}}, \bibinfo {author}
  {\bibfnamefont {D.}~\bibnamefont {Raffestin}}, \bibinfo {author}
  {\bibfnamefont {I.}~\bibnamefont {Thfoin-Lantuejoul}}, \bibinfo {author}
  {\bibfnamefont {B.}~\bibnamefont {Rosse}}, \bibinfo {author} {\bibfnamefont
  {C.}~\bibnamefont {Reverdin}}, \bibinfo {author} {\bibfnamefont
  {A.}~\bibnamefont {Duval}}, \bibinfo {author} {\bibfnamefont
  {F.}~\bibnamefont {Laniesse}}, \bibinfo {author} {\bibfnamefont
  {A.}~\bibnamefont {Chanc{\'{e}}}}, \bibinfo {author} {\bibfnamefont
  {D.}~\bibnamefont {Dubreuil}}, \bibinfo {author} {\bibfnamefont
  {B.}~\bibnamefont {Gastineau}}, \bibinfo {author} {\bibfnamefont {J.~C.}\
  \bibnamefont {Guillard}}, \bibinfo {author} {\bibfnamefont {F.}~\bibnamefont
  {Harrault}}, \bibinfo {author} {\bibfnamefont {D.}~\bibnamefont
  {Leb{\oe}uf}}, \bibinfo {author} {\bibfnamefont {J.-M.~L.}\ \bibnamefont
  {Ster}}, \bibinfo {author} {\bibfnamefont {C.}~\bibnamefont {P{\`{e}}s}},
  \bibinfo {author} {\bibfnamefont {J.-C.}\ \bibnamefont {Toussaint}}, \bibinfo
  {author} {\bibfnamefont {X.}~\bibnamefont {Leboeuf}}, \bibinfo {author}
  {\bibfnamefont {L.}~\bibnamefont {Lecherbourg}}, \bibinfo {author}
  {\bibfnamefont {C.~I.}\ \bibnamefont {Szabo}}, \bibinfo {author}
  {\bibfnamefont {J.-L.}\ \bibnamefont {Dubois}}, \ and\ \bibinfo {author}
  {\bibfnamefont {F.}~\bibnamefont {Lubrano-Lavaderci}},\ }\href {\doibase
  10.1088/0031-8949/2014/t161/014016} {\bibfield  {journal} {\bibinfo
  {journal} {Physica Scripta}\ }\textbf {\bibinfo {volume} {T161}},\ \bibinfo
  {pages} {014016} (\bibinfo {year} {2014})}\BibitemShut {NoStop}%
\bibitem [{\citenamefont {Crane}\ \emph {et~al.}(2010)\citenamefont {Crane},
  \citenamefont {Tietbohl}, \citenamefont {Arnold}, \citenamefont {Bliss},
  \citenamefont {Boley}, \citenamefont {Britten}, \citenamefont {Brunton},
  \citenamefont {Clark}, \citenamefont {Dawson}, \citenamefont {Fochs},
  \citenamefont {Hackel}, \citenamefont {Haefner}, \citenamefont {Halpin},
  \citenamefont {Heebner}, \citenamefont {Henesian}, \citenamefont {Hermann},
  \citenamefont {Hernandez}, \citenamefont {Kanz}, \citenamefont {McHale},
  \citenamefont {McLeod}, \citenamefont {Nguyen}, \citenamefont {Phan},
  \citenamefont {Rushford}, \citenamefont {Shaw}, \citenamefont {Shverdin},
  \citenamefont {Sigurdsson}, \citenamefont {Speck}, \citenamefont {Stolz},
  \citenamefont {Trummer}, \citenamefont {Wolfe}, \citenamefont {Wong},
  \citenamefont {Siders},\ and\ \citenamefont {Barty}}]{crane2010NIF_ARC}%
  \BibitemOpen
  \bibfield  {author} {\bibinfo {author} {\bibfnamefont {J.~K.}\ \bibnamefont
  {Crane}}, \bibinfo {author} {\bibfnamefont {G.}~\bibnamefont {Tietbohl}},
  \bibinfo {author} {\bibfnamefont {P.}~\bibnamefont {Arnold}}, \bibinfo
  {author} {\bibfnamefont {E.~S.}\ \bibnamefont {Bliss}}, \bibinfo {author}
  {\bibfnamefont {C.}~\bibnamefont {Boley}}, \bibinfo {author} {\bibfnamefont
  {G.}~\bibnamefont {Britten}}, \bibinfo {author} {\bibfnamefont
  {G.}~\bibnamefont {Brunton}}, \bibinfo {author} {\bibfnamefont
  {W.}~\bibnamefont {Clark}}, \bibinfo {author} {\bibfnamefont {J.~W.}\
  \bibnamefont {Dawson}}, \bibinfo {author} {\bibfnamefont {S.}~\bibnamefont
  {Fochs}}, \bibinfo {author} {\bibfnamefont {R.}~\bibnamefont {Hackel}},
  \bibinfo {author} {\bibfnamefont {C.}~\bibnamefont {Haefner}}, \bibinfo
  {author} {\bibfnamefont {J.}~\bibnamefont {Halpin}}, \bibinfo {author}
  {\bibfnamefont {J.}~\bibnamefont {Heebner}}, \bibinfo {author} {\bibfnamefont
  {M.}~\bibnamefont {Henesian}}, \bibinfo {author} {\bibfnamefont
  {M.}~\bibnamefont {Hermann}}, \bibinfo {author} {\bibfnamefont
  {J.}~\bibnamefont {Hernandez}}, \bibinfo {author} {\bibfnamefont
  {V.}~\bibnamefont {Kanz}}, \bibinfo {author} {\bibfnamefont {B.}~\bibnamefont
  {McHale}}, \bibinfo {author} {\bibfnamefont {J.~B.}\ \bibnamefont {McLeod}},
  \bibinfo {author} {\bibfnamefont {H.}~\bibnamefont {Nguyen}}, \bibinfo
  {author} {\bibfnamefont {H.}~\bibnamefont {Phan}}, \bibinfo {author}
  {\bibfnamefont {M.}~\bibnamefont {Rushford}}, \bibinfo {author}
  {\bibfnamefont {B.}~\bibnamefont {Shaw}}, \bibinfo {author} {\bibfnamefont
  {M.}~\bibnamefont {Shverdin}}, \bibinfo {author} {\bibfnamefont
  {R.}~\bibnamefont {Sigurdsson}}, \bibinfo {author} {\bibfnamefont
  {R.}~\bibnamefont {Speck}}, \bibinfo {author} {\bibfnamefont
  {C.}~\bibnamefont {Stolz}}, \bibinfo {author} {\bibfnamefont
  {D.}~\bibnamefont {Trummer}}, \bibinfo {author} {\bibfnamefont
  {J.}~\bibnamefont {Wolfe}}, \bibinfo {author} {\bibfnamefont {J.~N.}\
  \bibnamefont {Wong}}, \bibinfo {author} {\bibfnamefont {G.~C.}\ \bibnamefont
  {Siders}}, \ and\ \bibinfo {author} {\bibfnamefont {C.~P.~J.}\ \bibnamefont
  {Barty}},\ }\href {\doibase 10.1088/1742-6596/244/3/032003} {\bibfield
  {journal} {\bibinfo  {journal} {Journal of Physics: Conference Series}\
  }\textbf {\bibinfo {volume} {244}},\ \bibinfo {pages} {032003} (\bibinfo
  {year} {2010})}\BibitemShut {NoStop}%
\bibitem [{\citenamefont {Maywar}\ \emph {et~al.}(2008)\citenamefont {Maywar},
  \citenamefont {Kelly}, \citenamefont {Waxer}, \citenamefont {Morse},
  \citenamefont {Begishev}, \citenamefont {Bromage}, \citenamefont {Dorrer},
  \citenamefont {Edwards}, \citenamefont {Folnsbee}, \citenamefont
  {Guardalben}, \citenamefont {Jacobs}, \citenamefont {Jungquist},
  \citenamefont {Kessler}, \citenamefont {Kidder}, \citenamefont {Kruschwitz},
  \citenamefont {Loucks}, \citenamefont {Marciante}, \citenamefont {McCrory},
  \citenamefont {Meyerhofer}, \citenamefont {Okishev}, \citenamefont {Oliver},
  \citenamefont {Pien}, \citenamefont {Qiao}, \citenamefont {Puth},
  \citenamefont {Rigatti}, \citenamefont {Schmid}, \citenamefont {Shoup},
  \citenamefont {Stoeckl}, \citenamefont {Thorp},\ and\ \citenamefont
  {Zuegel}}]{maywar2008OMEGA_EP}%
  \BibitemOpen
  \bibfield  {author} {\bibinfo {author} {\bibfnamefont {D.~N.}\ \bibnamefont
  {Maywar}}, \bibinfo {author} {\bibfnamefont {J.~H.}\ \bibnamefont {Kelly}},
  \bibinfo {author} {\bibfnamefont {L.~J.}\ \bibnamefont {Waxer}}, \bibinfo
  {author} {\bibfnamefont {S.~F.~B.}\ \bibnamefont {Morse}}, \bibinfo {author}
  {\bibfnamefont {I.~A.}\ \bibnamefont {Begishev}}, \bibinfo {author}
  {\bibfnamefont {J.}~\bibnamefont {Bromage}}, \bibinfo {author} {\bibfnamefont
  {C.}~\bibnamefont {Dorrer}}, \bibinfo {author} {\bibfnamefont {J.~L.}\
  \bibnamefont {Edwards}}, \bibinfo {author} {\bibfnamefont {L.}~\bibnamefont
  {Folnsbee}}, \bibinfo {author} {\bibfnamefont {M.~J.}\ \bibnamefont
  {Guardalben}}, \bibinfo {author} {\bibfnamefont {S.~D.}\ \bibnamefont
  {Jacobs}}, \bibinfo {author} {\bibfnamefont {R.}~\bibnamefont {Jungquist}},
  \bibinfo {author} {\bibfnamefont {T.~J.}\ \bibnamefont {Kessler}}, \bibinfo
  {author} {\bibfnamefont {R.~W.}\ \bibnamefont {Kidder}}, \bibinfo {author}
  {\bibfnamefont {B.~E.}\ \bibnamefont {Kruschwitz}}, \bibinfo {author}
  {\bibfnamefont {S.~J.}\ \bibnamefont {Loucks}}, \bibinfo {author}
  {\bibfnamefont {J.~R.}\ \bibnamefont {Marciante}}, \bibinfo {author}
  {\bibfnamefont {R.~L.}\ \bibnamefont {McCrory}}, \bibinfo {author}
  {\bibfnamefont {D.~D.}\ \bibnamefont {Meyerhofer}}, \bibinfo {author}
  {\bibfnamefont {A.~V.}\ \bibnamefont {Okishev}}, \bibinfo {author}
  {\bibfnamefont {J.~B.}\ \bibnamefont {Oliver}}, \bibinfo {author}
  {\bibfnamefont {G.}~\bibnamefont {Pien}}, \bibinfo {author} {\bibfnamefont
  {J.}~\bibnamefont {Qiao}}, \bibinfo {author} {\bibfnamefont {J.}~\bibnamefont
  {Puth}}, \bibinfo {author} {\bibfnamefont {A.~L.}\ \bibnamefont {Rigatti}},
  \bibinfo {author} {\bibfnamefont {A.~W.}\ \bibnamefont {Schmid}}, \bibinfo
  {author} {\bibfnamefont {M.~J.}\ \bibnamefont {Shoup}}, \bibinfo {author}
  {\bibfnamefont {C.}~\bibnamefont {Stoeckl}}, \bibinfo {author} {\bibfnamefont
  {K.~A.}\ \bibnamefont {Thorp}}, \ and\ \bibinfo {author} {\bibfnamefont
  {J.~D.}\ \bibnamefont {Zuegel}},\ }\href {\doibase
  10.1088/1742-6596/112/3/032007} {\bibfield  {journal} {\bibinfo  {journal}
  {Journal of Physics: Conference Series}\ }\textbf {\bibinfo {volume} {112}},\
  \bibinfo {pages} {032007} (\bibinfo {year} {2008})}\BibitemShut {NoStop}%
\bibitem [{\citenamefont {Kemp}\ and\ \citenamefont
  {Divol}(2012)}]{kemp2012ps_2d}%
  \BibitemOpen
  \bibfield  {author} {\bibinfo {author} {\bibfnamefont {A.~J.}\ \bibnamefont
  {Kemp}}\ and\ \bibinfo {author} {\bibfnamefont {L.}~\bibnamefont {Divol}},\
  }\href {\doibase 10.1103/PhysRevLett.109.195005} {\bibfield  {journal}
  {\bibinfo  {journal} {Phys. Rev. Lett.}\ }\textbf {\bibinfo {volume} {109}},\
  \bibinfo {pages} {195005} (\bibinfo {year} {2012})}\BibitemShut {NoStop}%
\bibitem [{\citenamefont {Paradkar}, \citenamefont {Krasheninnikov},\ and\
  \citenamefont {Beg}(2012)}]{paradkar2012wellheating_theory}%
  \BibitemOpen
  \bibfield  {author} {\bibinfo {author} {\bibfnamefont {B.~S.}\ \bibnamefont
  {Paradkar}}, \bibinfo {author} {\bibfnamefont {S.~I.}\ \bibnamefont
  {Krasheninnikov}}, \ and\ \bibinfo {author} {\bibfnamefont {F.~N.}\
  \bibnamefont {Beg}},\ }\href {\doibase 10.1063/1.4731731} {\bibfield
  {journal} {\bibinfo  {journal} {Physics of Plasmas}\ }\textbf {\bibinfo
  {volume} {19}},\ \bibinfo {pages} {060703} (\bibinfo {year}
  {2012})}\BibitemShut {NoStop}%
\bibitem [{\citenamefont {Arber}\ \emph {et~al.}(2015)\citenamefont {Arber},
  \citenamefont {Bennett}, \citenamefont {Brady}, \citenamefont
  {Lawrence-Douglas}, \citenamefont {Ramsay}, \citenamefont {Sircombe},
  \citenamefont {Gillies}, \citenamefont {Evans}, \citenamefont {Schmitz},
  \citenamefont {Bell},\ and\ \citenamefont {Ridgers}}]{arber2015epoch}%
  \BibitemOpen
  \bibfield  {author} {\bibinfo {author} {\bibfnamefont {T.~D.}\ \bibnamefont
  {Arber}}, \bibinfo {author} {\bibfnamefont {K.}~\bibnamefont {Bennett}},
  \bibinfo {author} {\bibfnamefont {C.~S.}\ \bibnamefont {Brady}}, \bibinfo
  {author} {\bibfnamefont {A.}~\bibnamefont {Lawrence-Douglas}}, \bibinfo
  {author} {\bibfnamefont {M.~G.}\ \bibnamefont {Ramsay}}, \bibinfo {author}
  {\bibfnamefont {N.~J.}\ \bibnamefont {Sircombe}}, \bibinfo {author}
  {\bibfnamefont {P.}~\bibnamefont {Gillies}}, \bibinfo {author} {\bibfnamefont
  {R.~G.}\ \bibnamefont {Evans}}, \bibinfo {author} {\bibfnamefont
  {H.}~\bibnamefont {Schmitz}}, \bibinfo {author} {\bibfnamefont {A.~R.}\
  \bibnamefont {Bell}}, \ and\ \bibinfo {author} {\bibfnamefont {C.~P.}\
  \bibnamefont {Ridgers}},\ }\href {\doibase 10.1088/0741-3335/57/11/113001}
  {\bibfield  {journal} {\bibinfo  {journal} {Plasma Physics and Controlled
  Fusion}\ }\textbf {\bibinfo {volume} {57}},\ \bibinfo {pages} {113001}
  (\bibinfo {year} {2015})}\BibitemShut {NoStop}%
\bibitem [{\citenamefont {Arefiev}\ \emph {et~al.}(2015)\citenamefont
  {Arefiev}, \citenamefont {Cochran}, \citenamefont {Schumacher}, \citenamefont
  {Robinson},\ and\ \citenamefont {Chen}}]{arefiev2015resolution}%
  \BibitemOpen
  \bibfield  {author} {\bibinfo {author} {\bibfnamefont {A.~V.}\ \bibnamefont
  {Arefiev}}, \bibinfo {author} {\bibfnamefont {G.~E.}\ \bibnamefont
  {Cochran}}, \bibinfo {author} {\bibfnamefont {D.~W.}\ \bibnamefont
  {Schumacher}}, \bibinfo {author} {\bibfnamefont {A.~P.~L.}\ \bibnamefont
  {Robinson}}, \ and\ \bibinfo {author} {\bibfnamefont {G.}~\bibnamefont
  {Chen}},\ }\href {\doibase 10.1063/1.4905523} {\bibfield  {journal} {\bibinfo
   {journal} {Physics of Plasmas}\ }\textbf {\bibinfo {volume} {22}},\ \bibinfo
  {pages} {013103} (\bibinfo {year} {2015})}\BibitemShut {NoStop}%
\bibitem [{\citenamefont {Wilks}(1993)}]{wilks1993osc_surface}%
  \BibitemOpen
  \bibfield  {author} {\bibinfo {author} {\bibfnamefont {S.~C.}\ \bibnamefont
  {Wilks}},\ }\href {\doibase 10.1063/1.860697} {\bibfield  {journal} {\bibinfo
   {journal} {Physics of Fluids B: Plasma Physics}\ }\textbf {\bibinfo {volume}
  {5}},\ \bibinfo {pages} {2603} (\bibinfo {year} {1993})}\BibitemShut
  {NoStop}%
\bibitem [{\citenamefont {Bulanov}, \citenamefont {Naumova},\ and\
  \citenamefont {Pegoraro}(1994)}]{bulanov1994osc_surface}%
  \BibitemOpen
  \bibfield  {author} {\bibinfo {author} {\bibfnamefont {S.~V.}\ \bibnamefont
  {Bulanov}}, \bibinfo {author} {\bibfnamefont {N.~M.}\ \bibnamefont
  {Naumova}}, \ and\ \bibinfo {author} {\bibfnamefont {F.}~\bibnamefont
  {Pegoraro}},\ }\href {\doibase 10.1063/1.870766} {\bibfield  {journal}
  {\bibinfo  {journal} {Physics of Plasmas}\ }\textbf {\bibinfo {volume} {1}},\
  \bibinfo {pages} {745} (\bibinfo {year} {1994})}\BibitemShut {NoStop}%
\bibitem [{\citenamefont {Sheng}\ \emph {et~al.}(2004)\citenamefont {Sheng},
  \citenamefont {Mima}, \citenamefont {Zhang},\ and\ \citenamefont {Meyer-ter
  Vehn}}]{sheng2004stochastic_pic}%
  \BibitemOpen
  \bibfield  {author} {\bibinfo {author} {\bibfnamefont {Z.-M.}\ \bibnamefont
  {Sheng}}, \bibinfo {author} {\bibfnamefont {K.}~\bibnamefont {Mima}},
  \bibinfo {author} {\bibfnamefont {J.}~\bibnamefont {Zhang}}, \ and\ \bibinfo
  {author} {\bibfnamefont {J.}~\bibnamefont {Meyer-ter Vehn}},\ }\href
  {\doibase 10.1103/PhysRevE.69.016407} {\bibfield  {journal} {\bibinfo
  {journal} {Phys. Rev. E}\ }\textbf {\bibinfo {volume} {69}},\ \bibinfo
  {pages} {016407} (\bibinfo {year} {2004})}\BibitemShut {NoStop}%
\bibitem [{\citenamefont {Wu}\ \emph {et~al.}(2016)\citenamefont {Wu},
  \citenamefont {Krasheninnikov}, \citenamefont {Luan},\ and\ \citenamefont
  {Yu}}]{wu2016stoch_es}%
  \BibitemOpen
  \bibfield  {author} {\bibinfo {author} {\bibfnamefont {D.}~\bibnamefont
  {Wu}}, \bibinfo {author} {\bibfnamefont {S.}~\bibnamefont {Krasheninnikov}},
  \bibinfo {author} {\bibfnamefont {S.}~\bibnamefont {Luan}}, \ and\ \bibinfo
  {author} {\bibfnamefont {W.}~\bibnamefont {Yu}},\ }\href {\doibase
  10.1088/0029-5515/57/1/016007} {\bibfield  {journal} {\bibinfo  {journal}
  {Nuclear Fusion}\ }\textbf {\bibinfo {volume} {57}},\ \bibinfo {pages}
  {016007} (\bibinfo {year} {2016})}\BibitemShut {NoStop}%
\bibitem [{\citenamefont {Kruger}\ and\ \citenamefont
  {Bovyn}(1976)}]{kruger1976dla}%
  \BibitemOpen
  \bibfield  {author} {\bibinfo {author} {\bibfnamefont {J.}~\bibnamefont
  {Kruger}}\ and\ \bibinfo {author} {\bibfnamefont {M.}~\bibnamefont {Bovyn}},\
  }\href {\doibase 10.1088/0305-4470/9/11/008} {\bibfield  {journal} {\bibinfo
  {journal} {Journal of Physics A: Mathematical and General}\ }\textbf
  {\bibinfo {volume} {9}},\ \bibinfo {pages} {1841} (\bibinfo {year}
  {1976})}\BibitemShut {NoStop}%
\bibitem [{\citenamefont {Robinson}, \citenamefont {Arefiev},\ and\
  \citenamefont {Neely}(2013)}]{robinson2013dephasing}%
  \BibitemOpen
  \bibfield  {author} {\bibinfo {author} {\bibfnamefont {A.~P.~L.}\
  \bibnamefont {Robinson}}, \bibinfo {author} {\bibfnamefont {A.~V.}\
  \bibnamefont {Arefiev}}, \ and\ \bibinfo {author} {\bibfnamefont
  {D.}~\bibnamefont {Neely}},\ }\href {\doibase 10.1103/PhysRevLett.111.065002}
  {\bibfield  {journal} {\bibinfo  {journal} {Phys. Rev. Lett.}\ }\textbf
  {\bibinfo {volume} {111}},\ \bibinfo {pages} {065002} (\bibinfo {year}
  {2013})}\BibitemShut {NoStop}%
\bibitem [{\citenamefont {Robinson}\ \emph {et~al.}(2019)\citenamefont
  {Robinson}, \citenamefont {Tangtartharakul}, \citenamefont {Weichman},\ and\
  \citenamefont {Arefiev}}]{robinson2019solvers}%
  \BibitemOpen
  \bibfield  {author} {\bibinfo {author} {\bibfnamefont {A.~P.~L.}\
  \bibnamefont {Robinson}}, \bibinfo {author} {\bibfnamefont {K.}~\bibnamefont
  {Tangtartharakul}}, \bibinfo {author} {\bibfnamefont {K.}~\bibnamefont
  {Weichman}}, \ and\ \bibinfo {author} {\bibfnamefont {A.~V.}\ \bibnamefont
  {Arefiev}},\ }\href {\doibase 10.1063/1.5115993} {\bibfield  {journal}
  {\bibinfo  {journal} {Physics of Plasmas}\ }\textbf {\bibinfo {volume}
  {26}},\ \bibinfo {pages} {093110} (\bibinfo {year} {2019})}\BibitemShut
  {NoStop}%
\bibitem [{\citenamefont {Bochkarev}\ \emph {et~al.}(2019)\citenamefont
  {Bochkarev}, \citenamefont {d'Humi{\`{e}}res}, \citenamefont {Tikhonchuk},
  \citenamefont {Korneev},\ and\ \citenamefont
  {Bychenkov}}]{bochkarev2019stoch}%
  \BibitemOpen
  \bibfield  {author} {\bibinfo {author} {\bibfnamefont {S.~G.}\ \bibnamefont
  {Bochkarev}}, \bibinfo {author} {\bibfnamefont {E.}~\bibnamefont
  {d'Humi{\`{e}}res}}, \bibinfo {author} {\bibfnamefont {V.~T.}\ \bibnamefont
  {Tikhonchuk}}, \bibinfo {author} {\bibfnamefont {P.}~\bibnamefont {Korneev}},
  \ and\ \bibinfo {author} {\bibfnamefont {V.~Y.}\ \bibnamefont {Bychenkov}},\
  }\href {\doibase 10.1088/1361-6587/aaf7e1} {\bibfield  {journal} {\bibinfo
  {journal} {Plasma Physics and Controlled Fusion}\ }\textbf {\bibinfo {volume}
  {61}},\ \bibinfo {pages} {025015} (\bibinfo {year} {2019})}\BibitemShut
  {NoStop}%
\bibitem [{\citenamefont {Peebles}\ \emph {et~al.}(2017)\citenamefont
  {Peebles}, \citenamefont {Wei}, \citenamefont {Arefiev}, \citenamefont
  {McGuffey}, \citenamefont {Stephens}, \citenamefont {Theobald}, \citenamefont
  {Haberberger}, \citenamefont {Jarrott}, \citenamefont {Link}, \citenamefont
  {Chen}, \citenamefont {McLean}, \citenamefont {Sorokovikova}, \citenamefont
  {Krasheninnikov},\ and\ \citenamefont {Beg}}]{peebles2017preplasma}%
  \BibitemOpen
  \bibfield  {author} {\bibinfo {author} {\bibfnamefont {J.}~\bibnamefont
  {Peebles}}, \bibinfo {author} {\bibfnamefont {M.~S.}\ \bibnamefont {Wei}},
  \bibinfo {author} {\bibfnamefont {A.~V.}\ \bibnamefont {Arefiev}}, \bibinfo
  {author} {\bibfnamefont {C.}~\bibnamefont {McGuffey}}, \bibinfo {author}
  {\bibfnamefont {R.~B.}\ \bibnamefont {Stephens}}, \bibinfo {author}
  {\bibfnamefont {W.}~\bibnamefont {Theobald}}, \bibinfo {author}
  {\bibfnamefont {D.}~\bibnamefont {Haberberger}}, \bibinfo {author}
  {\bibfnamefont {L.~C.}\ \bibnamefont {Jarrott}}, \bibinfo {author}
  {\bibfnamefont {A.}~\bibnamefont {Link}}, \bibinfo {author} {\bibfnamefont
  {H.}~\bibnamefont {Chen}}, \bibinfo {author} {\bibfnamefont {H.~S.}\
  \bibnamefont {McLean}}, \bibinfo {author} {\bibfnamefont {A.}~\bibnamefont
  {Sorokovikova}}, \bibinfo {author} {\bibfnamefont {S.}~\bibnamefont
  {Krasheninnikov}}, \ and\ \bibinfo {author} {\bibfnamefont {F.~N.}\
  \bibnamefont {Beg}},\ }\href {\doibase 10.1088/1367-2630/aa5a21} {\bibfield
  {journal} {\bibinfo  {journal} {New Journal of Physics}\ }\textbf {\bibinfo
  {volume} {19}},\ \bibinfo {pages} {023008} (\bibinfo {year}
  {2017})}\BibitemShut {NoStop}%
\bibitem [{\citenamefont {Peebles}\ \emph {et~al.}(2018)\citenamefont
  {Peebles}, \citenamefont {Arefiev}, \citenamefont {Zhang}, \citenamefont
  {McGuffey}, \citenamefont {Spinks}, \citenamefont {Gordon}, \citenamefont
  {Gaul}, \citenamefont {Dyer}, \citenamefont {Martinez}, \citenamefont
  {Donovan}, \citenamefont {Ditmire}, \citenamefont {Park}, \citenamefont
  {Chen}, \citenamefont {McLean}, \citenamefont {Wei}, \citenamefont
  {Krasheninnikov},\ and\ \citenamefont {Beg}}]{peebles2018preplasma}%
  \BibitemOpen
  \bibfield  {author} {\bibinfo {author} {\bibfnamefont {J.}~\bibnamefont
  {Peebles}}, \bibinfo {author} {\bibfnamefont {A.~V.}\ \bibnamefont
  {Arefiev}}, \bibinfo {author} {\bibfnamefont {S.}~\bibnamefont {Zhang}},
  \bibinfo {author} {\bibfnamefont {C.}~\bibnamefont {McGuffey}}, \bibinfo
  {author} {\bibfnamefont {M.}~\bibnamefont {Spinks}}, \bibinfo {author}
  {\bibfnamefont {J.}~\bibnamefont {Gordon}}, \bibinfo {author} {\bibfnamefont
  {E.~W.}\ \bibnamefont {Gaul}}, \bibinfo {author} {\bibfnamefont
  {G.}~\bibnamefont {Dyer}}, \bibinfo {author} {\bibfnamefont {M.}~\bibnamefont
  {Martinez}}, \bibinfo {author} {\bibfnamefont {M.~E.}\ \bibnamefont
  {Donovan}}, \bibinfo {author} {\bibfnamefont {T.}~\bibnamefont {Ditmire}},
  \bibinfo {author} {\bibfnamefont {J.}~\bibnamefont {Park}}, \bibinfo {author}
  {\bibfnamefont {H.}~\bibnamefont {Chen}}, \bibinfo {author} {\bibfnamefont
  {H.~S.}\ \bibnamefont {McLean}}, \bibinfo {author} {\bibfnamefont {M.~S.}\
  \bibnamefont {Wei}}, \bibinfo {author} {\bibfnamefont {S.~I.}\ \bibnamefont
  {Krasheninnikov}}, \ and\ \bibinfo {author} {\bibfnamefont {F.~N.}\
  \bibnamefont {Beg}},\ }\href {\doibase 10.1103/PhysRevE.98.053202} {\bibfield
   {journal} {\bibinfo  {journal} {Phys. Rev. E}\ }\textbf {\bibinfo {volume}
  {98}},\ \bibinfo {pages} {053202} (\bibinfo {year} {2018})}\BibitemShut
  {NoStop}%
\bibitem [{\citenamefont {Chourasia}, \citenamefont {Nadeau},\ and\
  \citenamefont {Norman}(2017)}]{chourasia2017seedme}%
  \BibitemOpen
  \bibfield  {author} {\bibinfo {author} {\bibfnamefont {A.}~\bibnamefont
  {Chourasia}}, \bibinfo {author} {\bibfnamefont {D.}~\bibnamefont {Nadeau}}, \
  and\ \bibinfo {author} {\bibfnamefont {M.}~\bibnamefont {Norman}},\ }in\
  \href {\doibase 10.1145/3093338.3104153} {\emph {\bibinfo {booktitle}
  {Proceedings of the Practice and Experience in Advanced Research Computing
  2017 on Sustainability, Success and Impact}}},\ \bibinfo {series and number}
  {PEARC17}\ (\bibinfo  {publisher} {ACM},\ \bibinfo {address} {New York, NY,
  USA},\ \bibinfo {year} {2017})\ pp.\ \bibinfo {pages}
  {69:1--69:1}\BibitemShut {NoStop}%
\bibitem [{\citenamefont {Arefiev}\ \emph {et~al.}(2016)\citenamefont
  {Arefiev}, \citenamefont {Khudik}, \citenamefont {Robinson}, \citenamefont
  {Shvets}, \citenamefont {Willingale},\ and\ \citenamefont
  {Schollmeier}}]{arefiev2016vphi}%
  \BibitemOpen
  \bibfield  {author} {\bibinfo {author} {\bibfnamefont {A.~V.}\ \bibnamefont
  {Arefiev}}, \bibinfo {author} {\bibfnamefont {V.~N.}\ \bibnamefont {Khudik}},
  \bibinfo {author} {\bibfnamefont {A.~P.~L.}\ \bibnamefont {Robinson}},
  \bibinfo {author} {\bibfnamefont {G.}~\bibnamefont {Shvets}}, \bibinfo
  {author} {\bibfnamefont {L.}~\bibnamefont {Willingale}}, \ and\ \bibinfo
  {author} {\bibfnamefont {M.}~\bibnamefont {Schollmeier}},\ }\href {\doibase
  10.1063/1.4946024} {\bibfield  {journal} {\bibinfo  {journal} {Physics of
  Plasmas}\ }\textbf {\bibinfo {volume} {23}},\ \bibinfo {pages} {056704}
  (\bibinfo {year} {2016})}\BibitemShut {NoStop}%
\bibitem [{\citenamefont {Willingale}\ \emph {et~al.}(2018)\citenamefont
  {Willingale}, \citenamefont {Arefiev}, \citenamefont {Williams},
  \citenamefont {Chen}, \citenamefont {Dollar}, \citenamefont {Hazi},
  \citenamefont {Maksimchuk}, \citenamefont {Manuel}, \citenamefont {Marley},
  \citenamefont {Nazarov}, \citenamefont {Zhao},\ and\ \citenamefont
  {Zulick}}]{willingale2018vphi}%
  \BibitemOpen
  \bibfield  {author} {\bibinfo {author} {\bibfnamefont {L.}~\bibnamefont
  {Willingale}}, \bibinfo {author} {\bibfnamefont {A.~V.}\ \bibnamefont
  {Arefiev}}, \bibinfo {author} {\bibfnamefont {G.~J.}\ \bibnamefont
  {Williams}}, \bibinfo {author} {\bibfnamefont {H.}~\bibnamefont {Chen}},
  \bibinfo {author} {\bibfnamefont {F.}~\bibnamefont {Dollar}}, \bibinfo
  {author} {\bibfnamefont {A.~U.}\ \bibnamefont {Hazi}}, \bibinfo {author}
  {\bibfnamefont {A.}~\bibnamefont {Maksimchuk}}, \bibinfo {author}
  {\bibfnamefont {M.~J.-E.}\ \bibnamefont {Manuel}}, \bibinfo {author}
  {\bibfnamefont {E.}~\bibnamefont {Marley}}, \bibinfo {author} {\bibfnamefont
  {W.}~\bibnamefont {Nazarov}}, \bibinfo {author} {\bibfnamefont {T.~Z.}\
  \bibnamefont {Zhao}}, \ and\ \bibinfo {author} {\bibfnamefont
  {C.}~\bibnamefont {Zulick}},\ }\href {\doibase 10.1088/1367-2630/aae034}
  {\bibfield  {journal} {\bibinfo  {journal} {New Journal of Physics}\ }\textbf
  {\bibinfo {volume} {20}},\ \bibinfo {pages} {093024} (\bibinfo {year}
  {2018})}\BibitemShut {NoStop}%
\end{thebibliography}
%

\end{document}